\documentclass[aps, prd, twocolumn, showpacs, floatfix, letterpaper,
 nofootinbib, amsmath, amssymb, superscriptaddress]{revtex4-1}
\usepackage{graphicx}
\usepackage{epsfig}
\usepackage{bm}
\usepackage{amsfonts}
\usepackage{pgf}
\usepackage{color}
\usepackage[T1]{fontenc}
\usepackage[latin1]{inputenc}
\usepackage{graphicx}
\usepackage[english]{babel}
\usepackage{amsmath}
\usepackage{amssymb}
\usepackage{amsfonts}
\usepackage{makecell}
\usepackage{hyperref}
\usepackage[normalem]{ulem}
\usepackage{xspace}

\def\e{\mathrm{e}}
\newcommand{\dd}{\mathrm{d}}
\newcommand{\A}{\textbf{A}\xspace}
\newcommand{\B}{\textbf{B}\xspace}

\newcommand{\tzeta}{\tilde{\zeta}}
\newcommand{\thg}{\tilde{h}}

\newcommand{\ade}{{a_\mathrm{de}}}
\newcommand{\ab}{a_\mathrm{b}}
\newcommand{\alphab}{\alpha_\mathrm{b}}
\newcommand{\X}{\mathcal{X}}
\newcommand{\Xb}{\X_\mathrm{b}}

\newcommand{\GN}{G_{_\mathrm{N}}}

\newcommand{\Mp}{M_{_\mathrm{P}}}
\newcommand{\lP}{\ell_{_\mathrm{P}}}
\newcommand{\Lag}{\mathcal{L}}

\begin{document}

\title{Consistent Scalar and Tensor Perturbation Power Spectra in Single Fluid Matter Bounce with Dark Energy Era}

\author{Anna Paula Bacalhau}
\email{anna@cbpf.br}
\affiliation{COSMO -- Centro Brasileiro de Pesquisas
F\'isicas, Xavier Sigaud, 150, Urca, Rio de Janeiro, Brasil}

\author{Nelson Pinto-Neto}
\email{nelsonpn@cbpf.br}
\affiliation{COSMO -- Centro Brasileiro de Pesquisas
F\'isicas, Xavier Sigaud, 150, Urca, Rio de Janeiro, Brasil}

\author{Sandro Dias Pinto Vitenti}
\email{sandro@isoftware.com.br}
\affiliation{COSMO -- Centro Brasileiro de Pesquisas
	F\'isicas, Xavier Sigaud, 150, Urca, Rio de Janeiro, Brasil}
\affiliation{Center for Cosmology, Particle Physics and Phenomenology,
Institute of Mathematics and Physics, Louvain University, 2 Chemin
du Cyclotron, 1348 Louvain-la-Neuve, Belgium}

\begin{abstract}
We investigate cosmological scenarios containing one canonical scalar
field with an exponential potential in the context of bouncing models,
where the bounce happens due to quantum cosmological effects. The only
possible bouncing solutions in this scenario (discarding an infinitely
fine tuned exception) must have one and only one dark energy phase,
either occurring in the contracting era or in the expanding era. Hence,
these bounce solutions are necessarily asymmetric. Naturally, the more
convenient solution is the one where the dark
energy phase happens in the expanding era, in order to be a possible
explanation for the current accelerated expansion indicated by
cosmological observations. In this case, one has the picture of a
Universe undergoing a classical dust contraction from very large scales,
the initial repeller of the model, moving to a classical stiff matter
contraction near the singularity, which is avoided due to the quantum
bounce. The Universe is then launched to a dark energy era, after passing
through radiation and dust dominated phases, finally returning to the
dust expanding phase, the final attractor of the model. We calculate the
spectral indexes and amplitudes of scalar and tensor perturbations
numerically, considering the whole history of the model, including the
bounce phase itself, without making any approximation or using any
matching condition on the perturbations. As the background model is
necessarily dust dominated in the far past, the usual adiabatic vacuum
initial conditions can be easily imposed in this era. Hence, this is a
cosmological model where the presence of dark energy behavior in the
Universe does not turn problematic the usual vacuum initial conditions
prescription for cosmological perturbation in bouncing models. Scalar and
tensor perturbations end up being almost scale invariant, as expected.
The background parameters can be adjusted, without fine tunings, to yield
the observed amplitude for scalar perturbations, and also for the ratio
between tensor and scalar amplitudes, $r = T/S \lesssim 0.1$. The
amplification of scalar perturbations over tensor perturbations takes
place only around the bounce, due to quantum effects, and it would not
occur if General Relativity has remained valid throughout this phase.
Hence, this is a bouncing model where a single field induces not only an
expanding background dark energy phase, but also produces all observed
features of cosmological perturbations of quantum mechanical origin at
linear order.
\end{abstract}

\date{\today}
\maketitle

\section{Introduction}
\label{Intro}

Bouncing models have been proposed as cosmological scenarios without an
initial singularity. Instead, the Universe had at least a preceding
contracting phase from very large length scales, shrinking the space
until the scale factor reaches a minimum value in which some new physics
takes place, mainly related to gravity modifications at very small length
scales, halting the contraction and launching the Universe into the
expanding phase we are living in.

In the standard cosmological model, inflation is responsible for
exponentially increase the particle horizon after the Big Bang in order
to explain why regions, which are not in causal contact at the last
scattering surface, present a highly correlated temperature distribution,
as observed today in the Cosmic Microwave Background radiation (CMB).
Without inflation, these regions would be causally disconnected in a
purely Big Bang model. This puzzle is the so called horizon problem, and
it does not exist in bouncing models. Since the Universe had a very large
period of contraction in the past, there is no limit to the particle
horizon (if the fluids dominating the contracting phase satisfy the
strong energy condition). Another puzzle of a purely Big Bang scenario is
the flatness problem, i.e., considering a Friedmann metric in a expanding
phase, the spatial curvature dilutes slower than any other matter content
(assuming again the strong energy condition) of the model. Hence, unless
the spatial curvature is strongly fine tuned to zero initially, it would
quickly dominates the expansion afterwards. When inflation is added to
the Big Bang scenario, it turns out that the Universe is driven
dynamically to an almost flat hypersurface, avoiding the initial
curvature fine tuning problem. In the bounce scenario this issue is not
posed, since flat space-like hypersurfaces are dynamical attractors during
contraction \cite{Peter2008, Brandenberger2012a}.

Despite the fact that inflation has not yet any consensual fundamental
physics behind it, a simple slow-roll prescription for the inflation
scalar field is enough to solve the above mentioned puzzles, and to
amplify quantum vacuum fluctuations after the Big Bang, thus giving rise
to an almost scale invariant adiabatic power spectrum, in good agreement
with CMB observations~\cite{PlanckCollaboration2015}. It is a challenge
for bounce cosmologies to reproduce a competitive fit for the
observations, and many models have been scrutinized over the years with
this aim.

In what concerns the primordial phase of bounce cosmologies, it has been
shown that perturbations originated from quantum vacuum fluctuations
during a matter dominated contraction phase become almost scale
invariant~\cite{Wands1999, Finelli2002, Pinho2007} in the expanding
phase. Whether the linear perturbation theory remains valid for a
specific gauge choice through the bounce is a subtle question addressed
by many authors~\cite{Cartier2003, Martin2003a, Martin2004, Allen2004},
and finally clarified in Refs.\cite{Vitenti2012, Pinto-Neto2014},
confirming the validity of linear perturbation theory up to the expanding
phase. The background scenarios used to developed those investigations
are matter contractions driven either by a canonical scalar field with an
exponential potential, a K-essence scalar field representing a
hydrodynamical fluid, or a relativistic perfect fluid using Schutz
formalism~\cite{Schutz1970, Schutz1971}. This scenario with a contracting
phase dominated by a dust-like fluid is called matter bounce scenario.
This class of models is an interesting new approach to the Big
Bang/Inflation scenario~\cite{Brandenberger2012a, Allen2004, Peter2008,
Cai2012}, and they have been extensively studied over the past 15 years.
A weak point for any model including a contracting phase, where all the
matter content satisfy the dominant energy condition, is the presence of
Belinsky-Khalatnikov-Lifshitz (BKL) instabilities~\cite{Belinskii1970a,
Battefeld2015, Karouby2011, Brandenberger2016}, the fast growth of
anisotropies during the contraction. Some proposals inspired in the
Ekpyrotic Model~\cite{Khoury2001} address this problem by means of an
ad-hoc ekpyrotic type potential~\cite{Cai:2013vm}. It is not the aim of
this work to address this kind of issue, since it is possible to overcome
it in more complex scenarios without completely spoiling out a suitable
primordial power spectra. Note that any cosmological model, either
inflationary, bouncing, or any other, has a much more serious problem to
deal with: the large degree of initial homogeneity necessary to turn all
these models compatible with observations. This is largely more serious
than the BKL problem. Note also that once one assumes an initial
homogeneous and isotropic Universe, one can show for the models we are
considering that the shear perturbation will never overcome the
background degrees of freedom, even growing as fast as $a^{-6}$ in the
contracting phase~\cite{Pinto-Neto2014}, and hence the BKL problem is not
present once such assumption is made. For a discussion on that, see also
Ref.~\cite{Celani2017}.

In this paper we will carefully study the physical properties of
primordial quantum perturbations in a matter bounce realized by a
canonical scalar field with an exponential potential. Our starting point
are the results obtained in Refs.~\cite{Heard2002, Colistete2000}. In our
scheme, we will argue that, approaching the singularity during
contraction, quantum effects as calculated in Ref.~\cite{Colistete2000}
become relevant, and a bounce arises naturally in the context of the
canonical quantization of gravity, connecting the classical contracting
and expanding phases described in Ref.~\cite{Heard2002}. The known
cosmological solutions obtained through the de Broglie-Bohm (dBB)
formulation of quantum mechanics can be applied to this system since the
classical singularity takes place when the kinetic term dominates the
scalar field dynamics, and the potential becomes negligible, exactly as
in the model investigated in Ref.~\cite{Colistete2000}. Hence, one has
classical contracting and expanding phases connected by a quantum bounce.
Our background model avoids the need of a ghost scalar field and is
sustained by the fact that, in the regime where the curvature scale is
$10^2$ Planck length or larger, the canonical quantization we implement
is expected to be an effective limit of more fundamental theories of
quantum gravity. Finally, since the perturbations evolve through the
background quantum phase, we use the right action for the perturbations
when the background is not assumed to be classical~\cite{Falciano2013}.

Our dynamical system analysis shows that this scenario carries an
interesting feature: as we will see, if a bounce takes place in between
classical contracting and expanding phases, the scalar field presents an
effective transient dark energy-type Equation of State (EoS), either in
the past of the contracting phase or in the future of the expanding
phase. This addresses an aspect of bouncing cosmologies that has gained
increased attention~\cite{Zhang2007, Lehners2009, Jamil2011, Maier2011,
Brevik2014, Odintsov2016}: the role of the dark energy (DE) in the
contracting phase of bouncing models. In the expansion history probed by
current observations, the effects due to the existence of a DE component
can only be felt when the scale factor is of about 3-4 e-folds from the
last scattering surface~\cite{Dodelson2003}. Whether the DE is a
cosmological constant or a quintessence field, it should be present
during the contracting phase too. Thus, it may affect the contracting
phase evolution of perturbations sensible to scales around the same
Hubble radius as today. The presence of dark energy in the contracting
phase of bouncing models may turn problematic the imposition of vacuum
initial conditions for cosmological perturbations in the far past of such
models. For instance, if dark energy is a simple cosmological constant,
all modes will eventually become larger than the curvature scale in the
far past, and an adiabatic vacuum prescription becomes quite contrived,
see Ref.~\cite{Maier2011} for a discussion on this point. However, in the
case of a scalar field with exponential potential, which contains a
transient dark energy phase, the Universe will always be dust dominated
in the far past, and adiabatic vacuum initial conditions can be easily
imposed in this era, as usual. Hence, this is a situation where the
presence of dark energy does not turn problematic the usual initial
conditions prescription for cosmological perturbations in bouncing
models.

The bounce solutions obtained here are necessarily asymmetric, i.e., the
transient DE is effective either in the contracting phase or in the
expanding phase, but not in both. Here we will study the more realistic
solution where the dark energy phase happens in the expanding era,
connecting the bounce model with the current accelerated expansion phase.
Hence one has the picture of a Universe which realizes a dust contraction
from very large scales, the initial repeller of the model, moves to a
stiff matter contraction near the singularity and realizes a quantum
bounce that ejects the Universe in a stiff matter expanding phase. The
latter moves to a dark energy era, finally returning to the dust
expanding phase, the final attractor of the model. The other possibility,
DE in the contracting phase, is more academic, and we let it for a future
work. The background solutions are constructed numerically, matching the
classical and quantum eras in the phase where both have the similar
dynamics.

Note that it can already be found in the literature exact solutions of
the full Wheeler-DeWitt equation for a canonical scalar field with
exponential potential, which is not neglected in the quantum
phase~\cite{Colin2017}. These solutions were obtained without matchings.
However, as they have exactly the same physical features as the solutions
described above (classical behavior up to stiff matter domination, where
quantum effects begin to be important and the potential is negligible,
and one and only one DE energy phase all along), we preferred to adopt
the above matching procedure, in which the numerical calculations are
simpler to handle.

After the background construction, scalar and tensor perturbations are
calculated numerically, and the results are understood analytically.
Depending on the parameters of the background, they turn out to be almost
scale invariant, with the right observed amplitude for scalar
perturbations, and also for the ratio between tensor and scalar
amplitudes, $r = T/S \lesssim 0.1$. The amplification of scalar
perturbations over tensor perturbations takes place only around the
bounce, and we explicitly show that it happens due to the quantum effects
on the background model producing the bounce. There are many papers
pointing out the difficulties of producing this amplification of scalar
perturbations over tensor perturbations in the framework of General
Relativity (GR), see Refs.~\cite{Allen2004, Xue2013, Battarra2014}.
Indeed, the amplification we will present would not occur if GR has
remained valid all along the bounce. Hence, our result shows that when GR
is violated around the bounce, the influence of this phase on the
evolution of cosmological perturbations can be nontrivial, and must be
evaluated with care. These effects  provide a counter example to the usual case where the
perturbations are unaffected by the details of the bounce and their
amplitude are determined by the bounce depth (the ratio between the value
of the scale factor during the potential crossing and its value at the
bounce).

The paper will be divided as follows: in Sec.~\ref{sec_back}, based on
Ref.~\cite{Heard2002}, we summarize the classical minisuperspace model
and its full space of solutions. In Sec.~\ref{sec_quant}, we present the
quantum background near the singularity, as presented in
Ref.~\cite{Colistete2000}. The matching of the classical and quantum
solutions is explained in Sec.~\ref{sec_matc}, where we obtain the full
background model with reasonable observational properties.
Section~\ref{sec_pert} describes the equations of motion for the quantum
perturbations with suitable vacuum initial conditions, and we perform the
numerical calculations in Sec.~\ref{sec_sol_pert}, exhibiting our final
results for both scalar and tensor perturbations. We conclude in
Sec.~\ref{sec_conclusion} with discussions and perspectives for future
work.

In what follows we will consider $\hbar = c = 1$ and the reduced Planck
mass $\Mp \equiv 1 / \kappa \equiv 1 / \sqrt{8 \pi \GN}$. The metric
signature is $(+, -, -, -)$.

\section{Background}
\label{sec_back}

We consider a canonical scalar field $\phi$ whose Lagrangian density is
given by
\begin{equation}
\Lag = \sqrt{-g}\left[\nabla^\nu \phi \nabla_\nu \phi - V(\phi)\right].
\end{equation}
The potential $V(\phi)$ is chosen to be the exponential, i.e.,
\begin{equation}
\label{def_pot}
V(\phi) = V_0 \e^{-\lambda \kappa \phi},
\end{equation}
where the constant $V_0$ has units $\mathrm{mass}^4$ and $\lambda$ is dimensionless.

The exponential potential has vastly assisted cosmologists to address
puzzles of the standard model because of its rich
dynamics. In
Refs.~\cite{Chang2016, Chen2016, Danila2016, Forte2016, Granda2016,
Harko2016, Oikonomou2016} we have some heterogeneous collection of what
was published with exponential potential only in 2016. For the background
dynamics, we will use results from~\cite{Halliwell1987, Copeland1998,
Kolda2001, Heard2002, Allen2004}.

In a flat, homogeneous and isotropic Universe, the
Friedmann-Lam\^aitre-Robertson-Walker metric is
\begin{equation}
\label{metric}
\dd s^ 2 =  N^2(\tau)\dd \tau^2 - a(\tau)^2 (\dd x^2 + \dd y^2 + \dd z^2 ),
\end{equation}
where $N(\tau)$ is the lapse function and $a(\tau)$ is the scale factor.
The evolution of the scale factor in cosmic time ($N(\tau) = 1, \;\tau =
t$) is given by the Friedmann equation coupled to the Klein-Gordon
equation, respectively,
\begin{align}
\dot{a} &= a H, \\
\dot{H} &= - \frac{\kappa^2}{2} \dot{\phi}^2 \label{der_Friedmann}, \\
\ddot{\phi} &= - 3 H \dot{\phi} - \frac{\dd V }{ \dd \phi} \label{klein_g},
\end{align}
where the dot operator represents the derivative with respect to the
cosmic time $t$. The Hubble function, $H$, must satisfy the Friedmann
constraint
\begin{equation}
\label{Friedmann_con}
H^2 = \frac{\kappa^ 2}{3} \left[ \frac{\dot{\phi}^ 2}{2} + V(\phi) \right].
\end{equation}

The background dynamics can be made simpler through a choice of
dimensionless variables that allows us to rewrite the coupled second
order equations \eqref{der_Friedmann} and \eqref{klein_g} as a planar
system \cite{Wainwright2005}, i.e.,
\begin{equation}\label{mud1}
x = \frac{\kappa \dot{\phi}}{\sqrt{6}H}, \qquad
y = \frac{\kappa \sqrt{V}}{\sqrt{3}H}.
\end{equation}
In those new variables, the Friedmann constraint,
Eq.~\eqref{Friedmann_con}, and the effective EoS read,
\begin{equation}
x^2 + y^2 = 1, \qquad  w = 2x^2-1. \label{x_y_con}
\end{equation}
Applying the above definitions to the system of
Eqs.~\eqref{der_Friedmann} and \eqref{klein_g} leads to the planar
system:
\begin{align}
\frac{\dd x}{\dd \alpha} &= - 3 x (1-x^2)+ \lambda \sqrt{\frac{3}{2}}y^2, \label{sist1} \\
\frac{\dd y}{\dd \alpha} &= x y \left(3x-\lambda \sqrt{\frac{3}{2}}\right), \label{sist2}
\end{align}
where $\alpha \equiv \ln (a)$. This system is supplemented by the
equations
\begin{equation}\label{eq:sup}
\dot{\alpha} = H, \qquad \dot{H} = -3H^2x^2.
\end{equation}
The critical points are listed in Tab.~\ref{tab_crit}.

\begin{table}
\begin{tabular}{|c|c|c|}
\hline
$x$ & $y$ & $w$ \\
\hline
$-1$ & $0$ &  $1 $\\
\hline
$1$ & $0$  & $ 1$ \\
\hline
$\frac{\lambda }{\sqrt{6}}$ &$ -\sqrt{1-\frac{\lambda ^2}{6}}$ &  $\frac{1}{3} \left(\lambda ^2-3\right)$ \\
\hline
$\frac{\lambda }{\sqrt{6}}$ & $\sqrt{1-\frac{\lambda ^2}{6}}$  & $\frac{1}{3} \left(\lambda ^2-3\right)$ \\
\hline
\end{tabular}
\caption{Critical points of the planar system Eqs.~\eqref{sist1} and
\eqref{sist2}.}
\label{tab_crit}
\end{table}

Near the critical points where $w = 1$, the effective energy density of
the scalar field evolves close to $a^{-6}$, i.e., it behaves
approximately like a stiff-matter fluid. For those where the effective
EoS is $w = \frac{1}{3} \left(\lambda^2 - 3\right) $, it evolves as
$a^{-\lambda^2}$. The qualitative behavior of the system can be studied
with the tools described in~\cite{Wainwright2005, Coley1999,
Boehmer2014}, for a detailed analysis see~\cite{Heard2002, Copeland1998,
Kolda2001, Allen2004}.

\begin{figure}[ht]
\begin{center}
\includegraphics[width=0.5\textwidth]{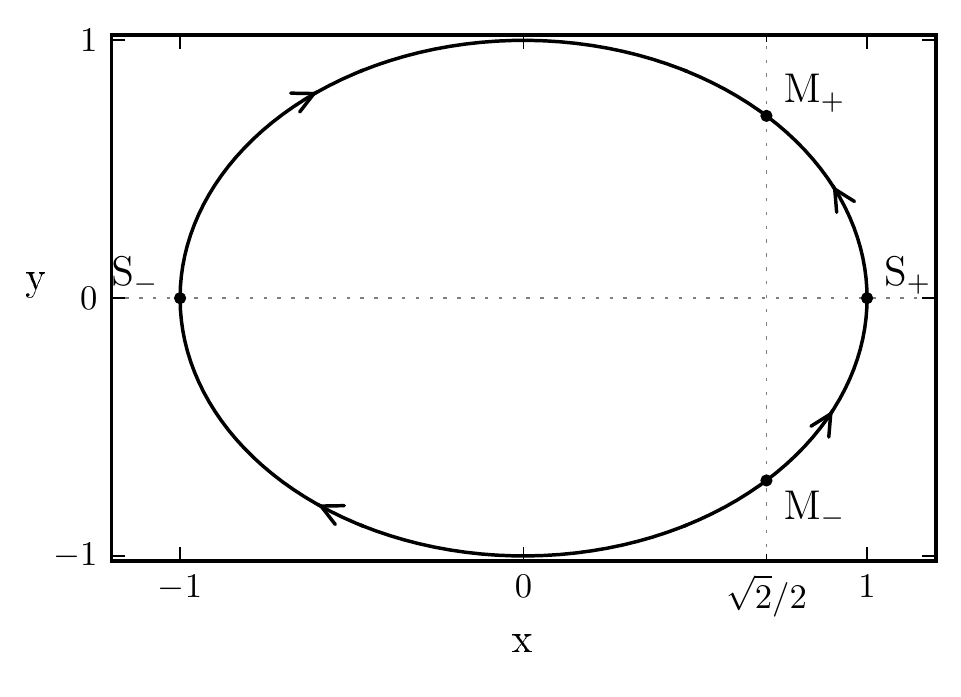}
\end{center}
\caption{Phase space for the planar system of Eqs.~\eqref{sist1} and
\eqref{sist2}. The critical points are indicated by $M_\pm$, for a scalar
field with a matter-type effective EoS, and $S_\pm$ for a stiff-matter
one. For $y < 0$ we have the contracting phase and for $y > 0$ the
expanding phase. Lower and upper quadrants are not physically connected,
because there is no classical mechanism that could drive a bounce between 
the contracting and expanding phases.}
\label{Fig:phase_full}
\end{figure}

In the contracting phase $H < 0 $, we can see by the definition of $y$,
Eq.~\eqref{mud1}, that $y < 0$. Note that $y$ is completely determined by
the value of $x$ through the constraint~\eqref{x_y_con} and the sign of
$H$. This phase is, therefore, tied to the lower quadrants of the phase
space, while the upper quadrants depict the expanding phase, see
Fig.~\ref{Fig:phase_full}. Using the Friedmann constraint
Eq.~\eqref{sist1} yields
\begin{equation}\label{sist1c}
\frac{\dd x}{\dd \alpha} = -3 \left(x-\frac{\lambda}{\sqrt{6}}\right) \left(1-x\right)\left(1+x\right).
\end{equation}
For $\lambda < \sqrt{6}$, the first two critical points, $x_{\pm{}c} =
\pm 1 $ and $y_{\pm{}c} = 0$, are unstable (repellers) during expansion and
stable (attractors) in the contraction phase, i.e.,
\begin{align*}
\left.\frac{\dd x}{\dd \alpha}\right\vert_{x=1-\epsilon} < 0, \qquad \left.\frac{\dd x}{\dd \alpha}\right\vert_{x=-1+\epsilon} > 0,
\end{align*}
for $0 < \epsilon\ll1$. For $\lambda > 0$, the critical point $x_{\lambda}
= {\lambda}/{\sqrt{6}}$ has the following behavior,
\begin{align*}
\left.\frac{\dd x}{\dd \alpha}\right\vert_{x=\frac{\lambda}{\sqrt{6}}+\epsilon} < 0, \qquad \left.\frac{\dd x}{\dd \alpha}\right\vert_{x=\frac{\lambda}{\sqrt{6}}-\epsilon} > 0.
\end{align*}
It is, therefore, an attractor during the expanding phase and an
a repeller in a contracting phase.

For the purpose of this work, we will choose $\lambda = \sqrt{3}$. As a
consequence, the scalar field behaves as a matter-fluid, $w = 0$, around
the critical point $x_\lambda$. Then, choosing the initial conditions for
$x$, as
\begin{equation}\label{xlambda}
x_\lambda = \frac{\lambda}{\sqrt{6}}\pm\epsilon,
\end{equation}
leads to a matter-fluid contracting phase. Since $x_\lambda$ is the only
attractor in the expanding phase, the model will always end in
matter-fluid dominated epoch.

In Fig.~\ref{Fig:phase_full} we have the phase space for the planar
system. The critical points in which the scalar field behaves as a
stiff-matter fluid are marked as $\mathrm{S_\pm}$ and the ones it behaves
as a matter-like fluid are $\mathrm{M_\pm}$. The contraction history goes
as follows: the Universe starts close to the critical point $M_-$ and,
depending on the choice of sign in Eq.~\eqref{xlambda}, arrives at the
stable point $S_+$ or $S_-$ for, respectively $+$ and $-$. The evolution
then ends up in a singularity (if no quantum effects are included). The
final EoS parameter is $w = 1$. Classically, there is no possible bounce
solution when the system arrives in the critical points $S_\pm$.

In the trajectories $M_- \rightarrow S_-$  and $S_{-} \rightarrow M_+$,
the Universe passes through a transient DE epoch, since
\begin{equation}\label{DEevol}
-1 < x < \frac{\lambda}{\sqrt{6}}\quad \Rightarrow\quad -1 \leq w < \frac{1}{3} \left(\lambda^2 - 3\right).
\end{equation}
Both possible expanding trajectories start in a stiff-matter epoch,
$S_\pm$, and  ends up in the matter epoch, $M_+$.

It will be useful to analyze what happens with $H$ and $\dot{\phi}$ when
the system is close to the critical points $S_\pm$. Around these points,
we get from Eq.~\eqref{eq:sup} that
\begin{equation}\label{evolxphi}
H \propto \e^{-3\alpha},
\end{equation}
as it is expected for a stiff-like fluid, which diverges when approaching
the singularity. Consequently, using Eq.~\eqref{mud1} we deduce that in
the neighborhood of these points the asymptotic behavior is
\begin{equation}
\lim_{x\to \pm 1}
\left\{\begin{array}{c l l}
H          &\to -\infty, \\
\phi       &\to \pm\frac{\sqrt{6}}{\kappa}\alpha &\to \mp \infty, \\
\dot{\phi} &\to \pm\frac{\sqrt{6}}{\kappa}H      &\to \mp \infty,
\end{array}\right.
\end{equation}
in the contracting phase, while in the expanding phase one gets
\begin{equation}
\lim_{x\to \pm 1}
\left\{\begin{array}{c l l}
H          &\to \infty, \\
\phi       &\to \pm\frac{\sqrt{6}}{\kappa}\alpha &\to \mp \infty, \\
\dot{\phi} &\to \pm\frac{\sqrt{6}}{\kappa}H      &\to \pm \infty.
\end{array}\right.
\end{equation}

\begin{figure}[ht]
\begin{center}
\includegraphics[width=0.5\textwidth]{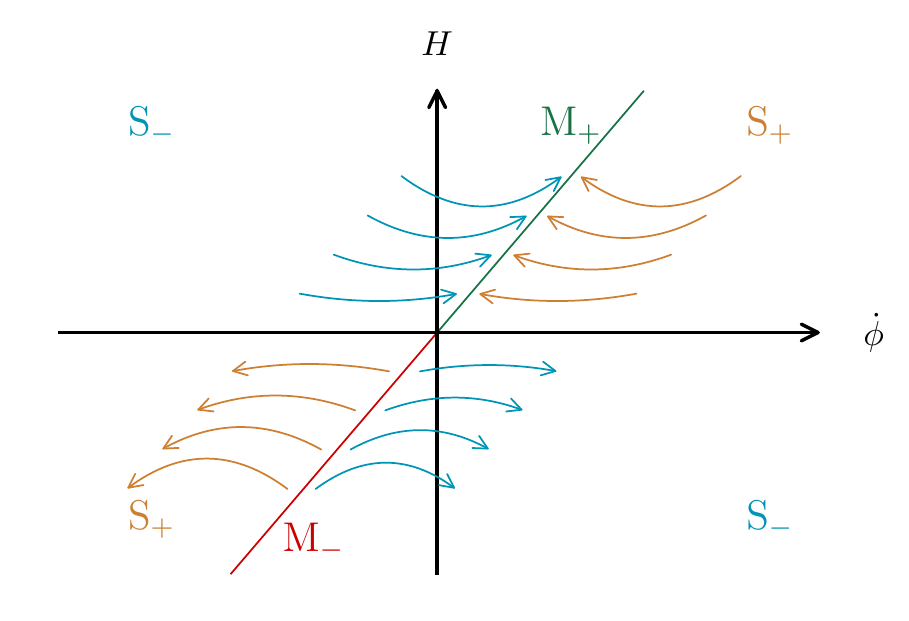}
\end{center}
\caption{This qualitative figure illustrates the behavior of the
solutions for $H$ and $\dot{\phi}$ close to the critical points of the
classical model Eqs.~\eqref{sist1} and \eqref{sist2}. In this figure we
chose $\kappa=\sqrt{6}$, hence the $M_{\pm}$ critical points are
represented by the line $H = \sqrt{2}\dot{\phi}$, while the $S_{\pm}$
critical points are represented by the lines $H = \pm \dot{\phi}$ for
$|H|\to \infty$ and $|\phi|\to\infty$. In a full quantized system, in
which the Universe bounces due to the quantum corrections close to the
Planck scale, the allowed phase space should connect the contraction
finishing in $S_{\pm}$ with the expansion beginning in $S_{\mp}$. }
\label{Fig:phase_quali}
\end{figure}

Figure~\ref{Fig:phase_quali} summarizes what we presented above
qualitatively. Nevertheless, this figure represents only the critical
points and flow resulting from the classical equations of motion. The
quantum dynamics takes place in the portion of phase space where the
Ricci scalar is close to the Planck length, near the singularities. As in
Fig.~\ref{Fig:phase_quali} we set $\kappa=\sqrt{6}$, the $S_{\pm}$
critical points representing the classical singularities are depicted in
this figure by the lines $H = \pm \dot{\phi}$ when $|H|\to \infty$ and
$|\dot{\phi}|\to\infty$. Hence, quantum effects can modify
Fig.~\ref{Fig:phase_quali} only around these regions, with a quantum
bounce connecting the regions around $S_\pm$ in the lower quadrants to
the regions around $S_\pm$ in the upper quadrants. However, trajectories
connecting the neighborhood of $S_+$ in the lower quadrant to the
neighborhood of $S_+$ in the upper quadrant, or similarly connecting the
neighborhoods of $S_-$ in both quadrants, necessarily cross the classical
$M_{\pm}$ line, $H = \sqrt{2} \dot{\phi}$. As we will see, in the de
Broglie-Bohm quantum theory, which we will use in this paper, velocity
fields yielding the quantum Bohmian trajectories are single valued
functions on phase space, as they arise from well defined functions on
this space (the phase's gradient from the corresponding Wheeler-DeWitt
wave equation's solution). Consequently, such trajectories cannot cross
each other. Hence, the only way to connect the almost singular
contracting and expanding behaviors without crossing the line $M_{\pm}$
is through a connection from the regions around $S_{\pm}$ to the regions
around $S_{\mp}$, in this reversed order, necessarily. Indeed, as we will
see below [see Eq.~\eqref{dphidt}], around the critical points $S_\pm$
the quantum bouncing trajectories have a well defined sign for
$\dot{\phi}$, which is determined completely from the model parameters. As
$H$ necessarily changes sign through a bounce, then these bouncing
trajectories can only connect the neighborhoods of $S_{\pm}$ to the ones
of $S_{\mp}$. Concluding, there is no solution in the complete phase
space where the Universe contracts in the direction of $S_-$ ($S_+$) and
expands from $S_-$ ($S_+$). A bounce is only possible in the phase space
of $H$ and $\dot{\phi}$ if it connects the contracting phase ending in
$S_+$ ($S_-$) to the expansion starting from $S_-$ ($S_+$).

\begin{figure}[ht]
\begin{center}
\includegraphics[width=0.5\textwidth]{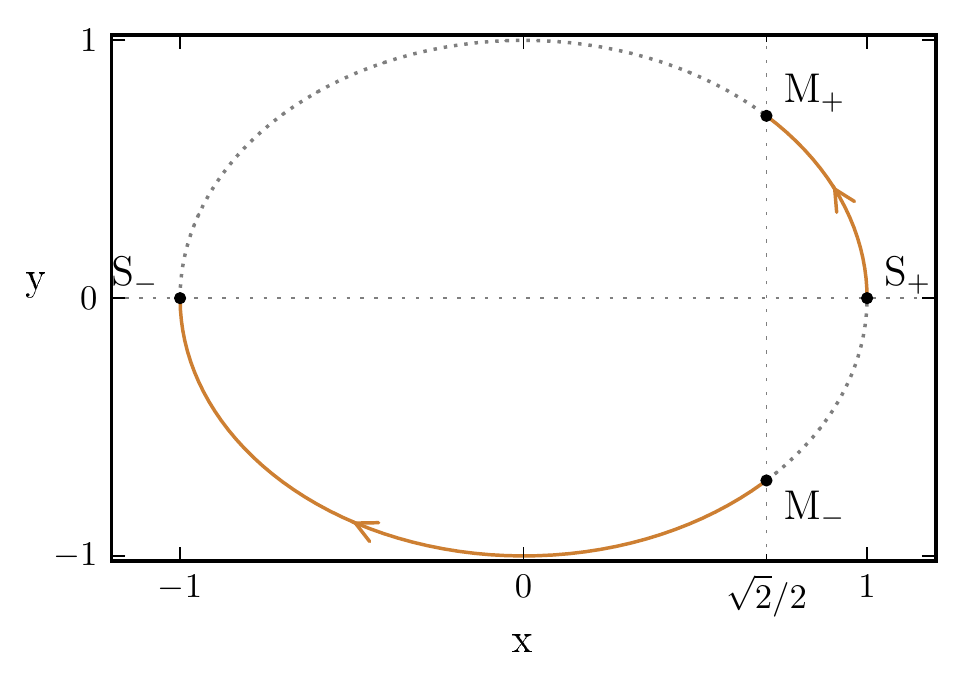}
\end{center}
\caption{Case \A: the scalar field has a DE-type equation of state during
contraction. By means of the quantum bounce, this system can not address
the DE in the future, since the matter attractor is reached before.}
\label{Fig:phase_2}
\end{figure}

\begin{figure}[ht]
\begin{center}
\includegraphics[width=0.5\textwidth]{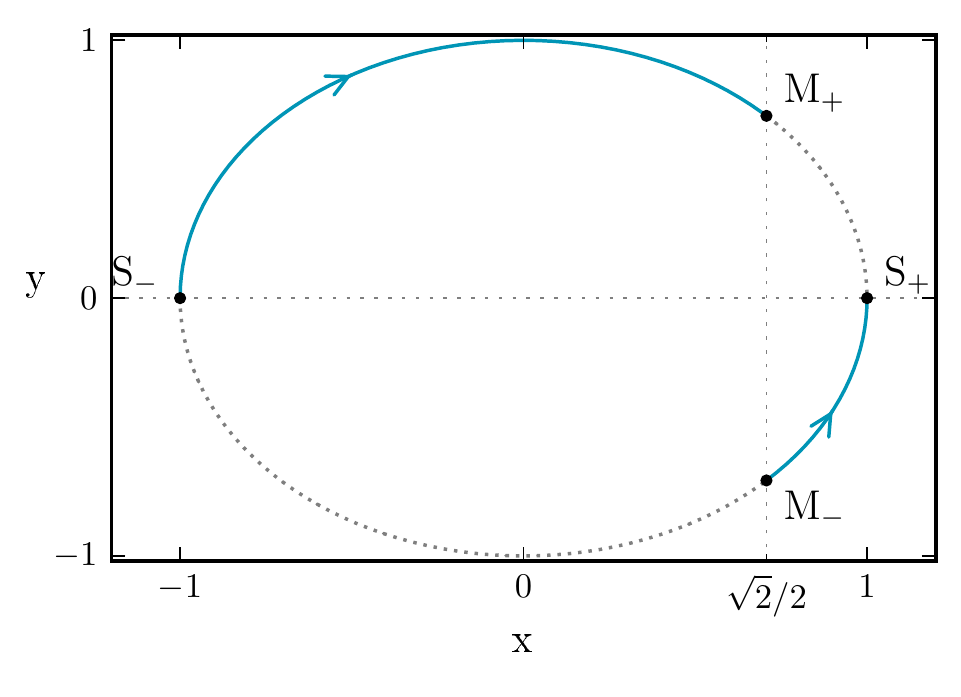}
\end{center}
\caption{Case \B: the contracting phase begins close to the unstable
point $M_-$, in which the scalar field has a dust-type EoS. After the
quantum bounce, the system emerges from $S_-$ and follows a DE phase
until reaches the future attractor $M_+$.} \label{Fig:phase_1}
\end{figure}

For the model in consideration, we have then two possible histories of
the Universe, depicted in Figs.~\ref{Fig:phase_2} and \ref{Fig:phase_1}.
In the case of Fig.~\ref{Fig:phase_2}, a contracting phase starts close
to $M_-$ (matter-fluid era), passes through a DE phase and ends in $S_-$
[as described in Eq.~\eqref{DEevol}], where the scalar field behaves as
an stiff-matter fluid. At this point, new physics takes place performing
a bounce and the Universe starts expanding from $S_+$ (scalar field as
stiff-matter) and ends in matter-like expansion in $M_+$. There is no DE
epoch in the expanding phase in this scenario.

The case of Fig.~\ref{Fig:phase_1} goes in the opposite direction.
Contraction happens from a matter epoch, $M_-$, to a stiff-matter one,
$S_+$. The new physics avoids the singularity and brings the Universe to
an expansion that starts in $S_-$ followed by a DE epoch, ending finally
in $M_+$, a matter-like epoch.

The two above-mentioned scenarios have the interesting feature of a
transient DE-like phase. For the sake of future reference along the
article, let us call the case with DE during contraction,
Fig.~\ref{Fig:phase_2}, case \A, and the one with DE during expansion,
Fig.~\ref{Fig:phase_1}, case \B.

Case \A shows the less compelling situation, where the background
performs a matter contraction followed by a DE epoch before the bounce,
but the expanding phase has no DE epoch. Previous works considering the
presence of DE during contraction used a ghost field to perform the
bounce. In such scenarios the phase space is very different from the one
presented here, and it can have a DE epoch in both contracting and
expanding phases~\cite{Heard2002, Cai2016}. Nevertheless, a complete and
rigorous calculation of the primordial power spectra in such scenarios
have not yet been performed.

Case \B is the one we will explore in this work. As we will see, the same
scalar field with exponential potential that realizes the matter bounce
scenario yielding an adiabatic scale invariant power spectrum also
produces a DE epoch in the expanding phase. Furthermore, there is no need
of an extra ghost potential nor any auxiliary scalar fields in order to
yield the physical conditions that produce the bounce.

As we mentioned earlier, there is no classical bounce in the previous
described backgrounds. Close to the attractor of the contracting phase,
$S_+$ in case \A and $S_-$ in case \B, $H \propto - a ^{-3}$, and when $a
\rightarrow0$ we reach a singularity. This happens because the kinetic
term dominates the Lagrangian of the scalar field and diverges, but in
these cases it has already been proved that quantum bounce solutions may
arise. In the next section we present the results from
\cite{Colistete2000}, and show how they can by applied to our case to
avoid the singularity.

\section{The quantum bounce}
\label{sec_quant}

Quantum cosmology is the field of research in which quantum theory is
applied to the Universe, and should have the standard cosmological model
as its classical limit. This interesting and challenging topic, not only
whitens fundamental problems of cosmology, as the singularity problem,
but also allows fundamental quantum mechanics to be tested at the
cosmological level~\cite{Calcagni2013, Peter2005, Martin2012,
Underwood2016, Benetti2016}.

The quantum description of gravity, besides facing many difficulties with
the non-renormalizable aspect of GR~\cite{Shomer2007}, also suffer from
fundamental conceptual issues in what concerns the application of a
quantum theory to the description of the whole Universe. In order to
construct a quantum theory for the Universe, the traditional Copenhagen
interpretation of quantum mechanics has to be replaced. Its main
limitation is the postulate of the collapse of the wave function
\cite{Wiltshire2000, Pinto-Neto2013}, where an outside classical system
is necessary to perform the collapse. This does not make sense if the
whole Universe is quantized.

The quantum theory which will replace the traditional Copenhagen point of
view must of course be able to reproduce the results of quantum
experiments already performed, but it must also dispense this exterior
classical world, or collapse recipes, in order to be applicable to
quantum cosmology. There are many proposals of quantum theories that
satisfy these criteria, and were already applied to Cosmology: the
consistent histories approach \cite{Griffiths1984, Zurek1990, Omnes1994,
Bom2014}, collapse models for the wave-function \cite{Pearle1976,
Ghirardi1987, Underwood2016, Benetti2016}, the many worlds interpretation
\cite{Everett1957, DeWitt1973, Page1999}, and the dBB quantum
theory~\cite{Bohm1993, Holland1993, Pinto-Neto2013}, which is the one we
will adopt here.

The canonical quantization of gravity obtained through the ADM formalism
\cite{Arnowitt2008, Dewitt1967}, which should be an effective limit of a
more fundamental theory, can be interpreted using the dBB formulation of
quantum mechanics. The dynamics of the wave function of the Universe is
given by the Wheeler-DeWitt equation from where, in this formulation, one
can obtain Bohmian trajectories with objective reality describing the
evolution of the whole system. In this approach, there is no need to
postulate any collapse of the wave function of the
Universe~\cite{Pinto-Neto2005, Pinto-Neto2013}.

Both models described in this paper depicted in Figs.~\ref{Fig:phase_2}
and \ref{Fig:phase_1} present the same feature: the end of the classical
contraction and the beginning of classical expansion happen when the
kinetic energy overcomes the scalar field potential $V(\phi)$. A system
consisting of a flat, homogeneous and isotropic space-time in the
presence of a scalar-field with a dominant kinetic term has already been
quantized. The Bohmian trajectories resulting from the Gaussian
superposition of plane wave functions led to bounce solutions. Details of
this construction can be found in \cite{Colistete2000}. We will summarize
their results in what follows.

In the case where the dynamics is dominated by the kinetic term, the
Hamiltonian for a scalar field in the metric~\eqref{metric} reads
\begin{equation}
\label{Ham}
H = N{\mathcal{H}} =  \frac{N\kappa^2}{12 V\e^{3 \alpha}} \left(- \Pi_{\alpha}^2 + \Pi_{\phi}^2\right),
\end{equation}
where $V$ is the volume of the conformal hypersurface, and from here
on we will be using the dimensionless scalar field
\begin{equation*}
\phi \rightarrow \frac{\kappa \phi}{\sqrt{6}}.
\end{equation*}
The associated momenta to the canonical variables $\alpha$ and $\phi$
are, respectively,
\begin{equation}
\Pi_{\alpha} = - \frac{6V}{N\kappa^2}\e^{3 \alpha} \dot{\alpha}, \qquad
\Pi_{\phi} = \frac{6V}{N\kappa^2}\e^{3 \alpha}\dot{\phi}.
\end{equation}

Finally, we choose the conformal hypersurface volume as $V = 4\pi\lP^3 /
3$, where $\lP \equiv \sqrt{\GN} = 1/(\sqrt{8\pi} \Mp)$ is the Planck
length. With this choice, the scale factor value has an absolute meaning,
i.e., when $a = 1$ the Universe has approximately the Planck volume.

Performing the Dirac quantization procedure, we can write the
Wheeler-DeWitt equation as
\begin{equation}\label{dirac_minisup}
{\hat{{\mathcal{H}}}}\Psi(\alpha,\phi) = 0 \Rightarrow
\left[- \frac{\partial^2}{\partial \alpha^2} + \frac{\partial ^2}{\partial \phi}\right] \Psi(\alpha,\phi) = 0.
\end{equation}
The general solution is
\begin{align}
\Psi (\alpha,\phi) &= F(\phi + \alpha) + G(\phi - \alpha) \nonumber\\ \label{psi_sep}
&\equiv \int \dd k \Big\{f(k) \exp\left[ik(\phi + \alpha)\right] + \\ \nonumber
&\qquad\qquad g(k) \exp\left[ik(\phi - \alpha)\right]\Big\},
\end{align}
where $f$ and $g$ are arbitrary functions of $k$.

Writing the wave-function in polar form, $\Psi = R \exp (i S)$, where $R$
is the amplitude and $S$ is the phase in units of $\hbar$, and
substituting into Eq.~\eqref{dirac_minisup} leads to the Hamilton-Jacobi
like equation for the phase $S$,
\begin{equation}\label{HJlike}
\left(\frac{\partial S}{\partial \alpha}\right)^2 - \left(\frac{\partial S}{\partial \phi}\right)^2 -
\frac{1}{R}\left(\frac{\partial^2 R}{\partial \alpha^2} - \frac{\partial ^2 R}{\partial \phi}\right) = 0.
\end{equation}
When the last term of Eq.~\eqref{HJlike}, the so called quantum
potential, is negligible with respect to the others, we get the usual
classical Hamilton-Jacobi equation for the minisuperspace model at hand.
Assuming the ontology of the trajectories $\alpha(t)$ and $\phi(t)$, this
classical limit suggests the imposition of the so called dBB guidance
relations in order to determine the trajectories, in correspondence to
the usual classical Hamilton-Jacobi theory, and they read, in cosmic time
$N=1$,
\begin{align}
\Pi_{\alpha} &= \frac{\partial S}{\partial \alpha} = - \lP\e^{3 \alpha} \dot{\alpha}
\label{guid_alpha}, \\
\Pi_{\phi} &= \frac{\partial S}{\partial \phi} =  \lP\e^{3 \alpha} \dot{\phi} .
\label{guid_phi}
\end{align}
When the quantum potential is not negligible in Eq.~\eqref{HJlike}, these
Bohmian quantum trajectories may be different from the classical ones,
and may present a bounce.

In Ref.~\cite{Colistete2000} it was chosen the simple appealing
prescription of a Gaussian superposition of plane waves in
Eq.~\eqref{psi_sep} given by
\begin{equation}
f(k) = g(k) = \exp \left[\frac{- (k - d)^2}{\sigma^2}\right].
\end{equation}
Calculating the phase $S$ of the aforementioned solution, and
substituting it into the guidance relations \eqref{guid_alpha} and
\eqref{guid_phi}, we find the planar system that describes the Bohmian
trajectories
\begin{align}
\label{dalphadt}
\lP\dot{\alpha} &= \frac{\phi \sigma ^2 \sin (2d\alpha) +
2d\sinh (\sigma ^2 \alpha  \phi)}
{2\e^{3\alpha}\left[ \cos (2d\alpha) +
\cosh (\sigma ^2 \alpha \phi)\right]}, \\
\label{dphidt}
\lP\dot{\phi} &= \frac{-\alpha \sigma ^2 \sin
(2d\alpha) +2d\cos(2d\alpha) + 2d\cosh (\sigma ^2 \alpha \phi)}
{2\e^{3\alpha}\left[ \cos (2d\alpha) +
\cosh (\sigma ^2 \alpha \phi)\right]}.
\end{align}
When solving the equations above we have a time definition in units of Planck
time (essentially putting $\lP = 1$ in the above equations). However,
since the scales of interest for the perturbations are those around the
Hubble radius today, $R_H \equiv 1/H_0$ (here we adopt the current value
$H_0 = 67.8 \pm 0.9\; \mathrm{km}\;\mathrm{s}^{-1}\mathrm{Mpc}^{-1}$
\cite{PlanckCollaboration2015}), we convert back using the factor $R_H /
\lP$ when matching with the classical solution.

\begin{figure}[ht]
\begin{center}
\includegraphics[width=0.5\textwidth]{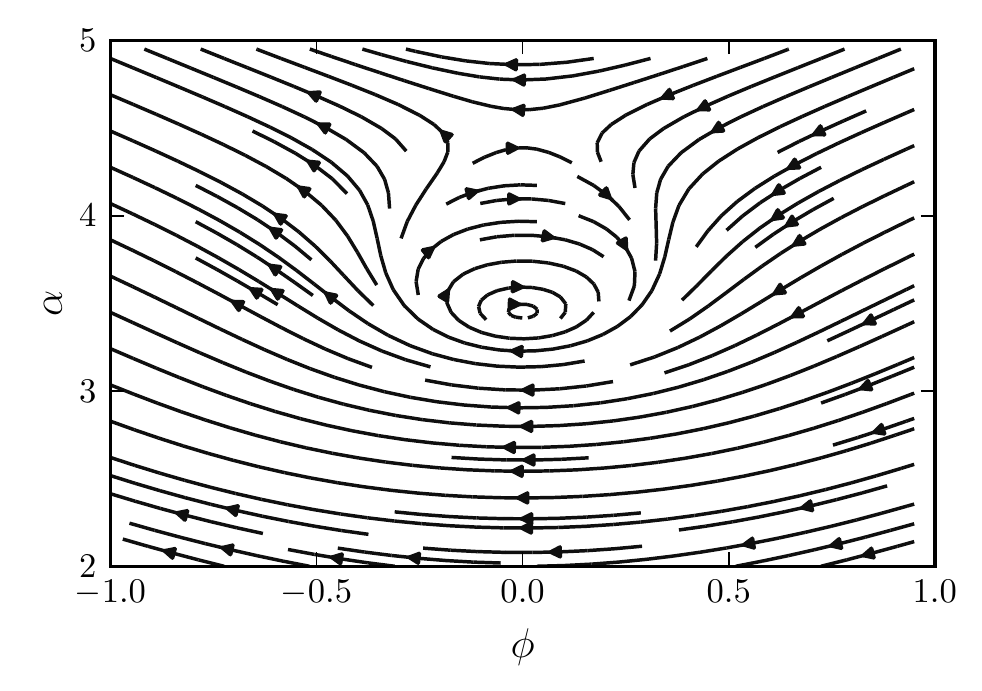}
\end{center}
\caption{Phase space for the system of Eqs.\eqref{dalphadt} and
\eqref{dphidt} for $d =-1$ and $\sigma = 1$. We can notice bouncing and
cyclic solutions. The bounces in the figure correspond to case \B, where
$\dot{\phi}<0$, and it connects regions around $S_+$ in the contracting
phase with regions around $S_-$ in the expanding phase.}
\label{Fig:phase_quantum}
\end{figure}

In Fig.~\ref{Fig:phase_quantum} we have the phase space for
Eqs.~\eqref{dalphadt} and \eqref{dphidt}. We can notice the presence of
bounce and cyclic solutions. It is easy to calculate the nodes and the
centers. They happen all along the line $\phi =0$: the nodes for $ d
\alpha = (2n +1)\pi / 2$ and the centers for  $\sigma^2 \alpha / 2 d =
\cot(d \alpha)$.

The classical limits of Eqs.~\eqref{dalphadt} and \eqref{dphidt} are
obtained for large $\alpha$, when the hyperbolic function dominates. From
the definition of $x$ in that limit, it is straightforward to obtain the
relations
\begin{align}
 x &\approx \coth \left(\sigma^2 \alpha \phi\right),
\label{cl_lim_x_qt}  \\
 \frac{H}{H_0} &\approx \frac{R_{H}}{\lP} \frac{d \e^{-3 \alpha}}{x}
\label{cl_lim_H_qt}, \\
\lP\dot{\phi} &\approx d\e^{-3\alpha}.
\end{align}
These equations imply that $\phi$ and $x$ have the same sign, and
$\dot{\phi}$ the same sign as $d$ in the classical limit. This means that
in case \A, since its contraction ends in $x \to -1$,
Eq.~\eqref{cl_lim_H_qt} is satisfied only if $d > 0$. Similarly, case \B
requires $d < 0$. This result is consistent with our discussion in
Sec.~\ref{sec_back}. In case \A, the quantum dynamics starts with $x
\approx -1$ ($\phi \ll -1$), ending in $x \approx 1$ ($\phi \gg 1$). The
opposite happens in case \B, i.e., our bouncing dynamics always connects
the classical critical points $S_-$ ($S_+$) with $S_+$ ($S_-$). In
practice, the sign of $d$ determines which case is being evolved.

\section{Matching of background}
\label{sec_matc}

In the previous section we presented the quantum corrections to the
system when the kinetic term of the scalar field dominates, yielding a
bounce. In order to construct a complete background, we should be able to
match the solutions from the classical evolution, described in
Sec.~\ref{sec_back}  with the quantum solution from the system of
Eqs.~\eqref{dalphadt} and \eqref{dphidt}.

In a complete formulation of this problem, the designation of
``classical'' and ``quantum'' solutions should not be taken strictly. In
the  complete dBB formulation, we have the Bohmian trajectories that
accounts for both regimes. In our hybrid background, we make this
distinction only to emphasize the fact that we do not have a complete
Bohmian trajectory and to make explicit which equations of motion are
being used. The full quantum description of this system can be found in
Ref.~\cite{Colin2017}, where full Bohmian bounce solutions are exhibited.
However, to perform the calculation of cosmological perturbations around
some full Bohmian trajectory is quite cumbersome, hence we prefer to
adopt this simpler method of matching classical to quantum solutions,
with no loss of relevant information.

The nomenclature in what follows may be tricky, and in order to avoid
confusion we will adopt the expressions \textit{quantum/classical
solutions}, \textit{regimes} or \textit{branches} to distinguish the
dynamics described by Eqs.~\eqref{dalphadt} and \eqref{dphidt} (quantum)
from the one determined by Eqs.~\eqref{sist1} and \eqref{sist2}
(classical). To make reference to the period at which the quantum potential is 
relevant, we will adopt \textit{quantum phase} in opposition
to \textit{classical phase}, in which the quantum potential is
irrelevant.

The complete background solution has three branches. The first one is the
classical contraction that starts with $x \approx 1 / \sqrt{2}$ and ends
in $x \rightarrow \pm 1$. The second branch is the quantum background
that starts in $x \approx \pm 1$ and bounces to $x \to \mp 1$. The
third branch, the classical expansion, starts with $x \approx \mp 1$
and ends with $x \rightarrow 1/\sqrt{2}$. The lower signs stand for case
\A while the upper signs for the case \B. The matching is performed
guaranteeing continuity of the solutions up to the first derivative at
the time when they move from the quantum to the classical regimes. This
happens when the quantum solutions reach their classical limit given in
Eqs.~\eqref{cl_lim_x_qt} and \eqref{cl_lim_H_qt}.

However, the classical limit of the quantum regime happens when $x = \pm
1 $, which is a critical point of the classical equations,
Eqs.~\eqref{sist1} and \eqref{sist2}. To start the classical phase
exactly in a critical point means that the Universe would be stuck in the
stiff-matter epoch. Nonetheless, the initial classical epoch is unstable,
hence we always start it with a small shift around the critical points.
We will parametrize $x$ in the proximity of the stiff critical points by
$x = \pm (1 - \epsilon_\pm)$, $0 < \epsilon_\pm \ll 1$. In the matching
point, $\epsilon_\pm$ should be small enough to justify the classical
limit of the quantum regime.

If we had a complete Bohmian trajectory, it would only be necessary to
set initial conditions in the far past. For instance, we would give an
initial $x_{\lambda}$ close to the unstable point $M_{-}$. The
proximity to the critical point and from which side of the critical point
it begins dictates the duration of the matter contraction and selects
between cases \A and \B (which must be compatible with the choice of sign
for $d$). We would also give an initial scale factor, $a_{\mathrm{ini}}$,
and a Hubble constant, $H_{\mathrm{ini}}$, ($\phi_{\mathrm{ini}}$ and
$\dot{\phi}_{\mathrm{ini}}$ are constrained by the value of
$x_{\mathrm{ini}}$ and $V_0$ through the Friedmann equation). With that
all set, the Bohmian trajectories would handle the whole evolution until
the expanding phase. Naturally, parameters as the minimum scale factor at
the bounce, the energy scale of the DE epoch, and the duration of the
quantum bounce would be obtainable from the model parameter $V_0$, the
system wave-function and the initial conditions.

Because of our matching procedure, things are not so simple. We have not
only the choice of initial conditions and the quantum bounce parameters,
$d$ and $\sigma$ (extracted from the wave-function), but also two new
variables, namely, the contraction and expansion matching parameters,
denoted by $\epsilon_c$ and $\epsilon_e$, respectively. In what follows,
we will rewrite these two variables in terms of
new parameters which also controls the number of e-folds between the
bounce and a given Hubble scale.

In order to connect the classical solution parameters at the matching point with its
quantum evolution, we integrate Eqs.~\eqref{eq:sup} and \eqref{sist1c}
analytically. However, it is not possible to write explicit functions for
$x(\alpha)$ and $H(\alpha)$. The best we can do is to obtain implicit
functions which, apart from two integration constants, read
\begin{align}
 3\alpha &= - \sqrt{2} \tanh^{-1}(x) - \ln \left[
\frac{\left(\frac{1}{\sqrt{2}} - x\right)^2}{1-x^2}\right] + \mathrm{cte}, \\
\ln{}H &= \sqrt{2} \tanh^{-1}(x) +\ \ln \left(\frac{ \frac{1}{\sqrt{2}} - x
}{1-x^2}\right) + \mathrm{cte} \label{sol_H_x}.
\end{align}

We begin by recasting these solutions in a more convenient form
\begin{align}
\left(\frac{a}{{\bar{a}}_0}\right)^6 \left(\frac{H}{H_0}\right)^2 &=
\frac{C_1}{\left( \frac{1}{\sqrt{2}}-x\right)^2},
\label{C1_sol_cl}\\
\left(\frac{a}{{\bar{a}}_0}\right)^3 &= \frac{C_2
\left(1-x\right)^{\gamma_{+}}\left(1+x\right)^{\gamma_{-}}}{\left(
\frac{1}{\sqrt{2}}-x\right)^2},\label{C2_sol_cl}
\end{align}
where  $\gamma_{\pm} \equiv 1 \pm \frac{1}{\sqrt{2}}$,  and $C_1$ and
$C_2$ are constants. We introduced $H_0$, the Hubble parameter today, and
${\bar{a}}_0$ for mere convenience. Note that these constants can be
absorbed in $C_1$ and $C_2$, and they do not represent any additional freedom
of the system. We can calculate the number
of e-folds between the critical points $S_\pm$ and $M_\pm$. Expanding
Eq.~\eqref{C2_sol_cl} around $x = \pm(1-\epsilon_\pm)$ and $x
=(1/\sqrt{2}\pm\epsilon_{\lambda})$ at leading order, yields,
respectively
\begin{align}\label{aapm}
\left(\frac{a_\pm}{{\bar{a}}_0}\right)^3 &\approx \frac{C_2
2^{\gamma_\mp}\epsilon_\pm^{\gamma_{\pm}}}{\gamma_\mp^2}, \\ \label{aalambda}
\left(\frac{a_\lambda}{{\bar{a}}_0}\right)^3 &\approx \frac{C_2
\gamma_-^{\gamma_{+}}\gamma_+^{\gamma_{-}}}{\epsilon_\lambda^2},
\end{align}
where we note that, at leading order, the second expression does not
depend from which side we approach $M_{\pm}$. From their ratio we get
\begin{equation}
\label{redshift1}
\left(\frac{a_\pm}{a_\lambda}\right)^3 \approx \frac{2^{\gamma_\pm}}{\gamma_-^{\gamma_{+}}
\gamma_+^{\gamma_{-}}\gamma_\mp^2}\epsilon_\pm^{\gamma_{\pm}}\epsilon_\lambda^2.
\end{equation}
Imposing that we must be close enough to the critical points, i.e.,
$\epsilon_i < 10^{-4}$ (for $i = \pm,\;\lambda$), Eq.~\eqref{redshift1}
leads to
\begin{equation*}
\frac{a_+}{a_\lambda} \lesssim \e^{-10}, \qquad \frac{a_-}{a_\lambda} \lesssim \e^{-6}.
\end{equation*}
This means that the trajectories $M_- \to S_-$ or $S_- \to M_+$, depicted
in the lower quadrants of Fig.~\ref{Fig:phase_2} and the upper quadrants
of Fig.~\ref{Fig:phase_1}, respectively, must have a minimum of $6$
e-folds. The remaining trajectories ($M_- \to S_+$ and $S_+ \to M_+$)
must have a minimum of $10$ e-folds.

Analogously, we can also calculate the number of e-folds until the DE phase ($x = 0$), yielding
\begin{equation*}
\left(\frac{\ade}{{\bar{a}}_0}\right)^3 \approx 2C_2, \quad \left(\frac{a_-}{\ade}\right)^3 \approx \frac{2^{\gamma_+}\epsilon_-^{\gamma_{-}}}{2\gamma_+^2}, \quad \frac{a_-}{\ade} \lesssim \e^{-1}.
\end{equation*}

The background will be constructed numerically from the quantum bounce to
the classical contracting and expanding phases. As the solutions are
necessarily asymmetric, the matching between the quantum and classical
regimes can be arranged in two possible ways, depending on whether we
want to write $\epsilon_c$ and $\epsilon_e$ in terms of the number of
e-folds between the bounce and a point in the matter-fluid domination
or the DE phase.
Let us now describe this construction in the following sub-sections.

\subsection{Initial conditions at the bounce}

Our numerical calculation consists in solving the background, starting
from the bounce and evolving to the expanding and contracting phases. In
order to accomplish this, we start the calculation around the bounce with
slightly positive (negative) time for the expansion (contraction) phase.
A convenient time variable is $\tau$ defined by
\begin{equation}\label{del_tau}
\alpha = \alphab + \frac{\tau^2}{2},
\end{equation}
which leads to $\dd \tau/ \dd t = H / \tau$ and $\dd \alpha = \tau
\dd\tau$. We have already included in this choice of time variable the
initial condition for $\alpha$, i.e., we always set the bounce at $\tau =
0$. We can see from Eq.~\eqref{dalphadt} that the initial value
$\alpha(t_0) = 0$ induces a trivial solution $\alpha(t) = 0$. As
trajectories in the $(\alpha , \phi)$ plane cannot cross, this implies
that $\alpha$ cannot chance sign along the possible trajectories. Hence,
the choice of time above select the positive branch of the phase space
$(\alpha , \phi)$, Fig. \ref{Fig:phase_quantum}. For a single bounce, $\alpha$ 
attains its smallest
value at the bounce, which provides the last justification for our
parametrization~\eqref{del_tau}.

We need now the initial condition for the field $\phi$. Examining
Eq.~\eqref{dalphadt}, we realize that the bounce can only take place when
$\phi = 0$. Indeed, the denominator is always positive, and it diverges
to the classical limit, where no bounce is possible. Hence the necessary
condition for the bounce $\dot{\alpha}=0$ can occur if and only if the
numerator is zero. If there were a root of the numerator of
Eq.~\eqref{dalphadt} different from the trivial one $\phi = 0$, it would
satisfy
\begin{equation*}
\frac{\sinh(A)}{A} = -\frac{\sin(B)}{B},
\end{equation*}
where we have defined $A = \sigma^2 \alpha  \phi$ and $B = 2d\alpha$,
both different from zero, by assumption. However, this equation cannot be
solved for any real $A$ and $B$. Therefore, for the quantum system,
we will always use the only possible initial conditions $\alpha(0) =
\alphab$ and $\phi(0) = 0$.

Expanding Eqs.~\eqref{dalphadt} and Eq.~\eqref{dphidt} about the bounce,
we get the leading order approximations
\begin{align}
\frac{\dd \tau}{\dd t_Q} = \frac{\phi}{\tau}D_1, \\
\frac{\dd \phi}{\dd t_Q} = D_2,
\end{align}
where we rewrote the equation in terms of $\tau$ and the convenient
dimensionless time variable $\e^{3\alpha}\lP\dd t_Q = \dd t$. The two
constants $D_1$ and $D_2$ are
\begin{align*}
D_1 &= \frac{\sigma^2\left[\sin(2d\alphab)+2d\alphab\right]}{2\left[2\cos(2d\alphab) + 1\right]}, \\
D_2 &= \frac{-\alphab \sigma^2 \sin(2d\alphab) + 2d\cos(2d\alphab)+2d}{2\left[2\cos(2d\alphab) + 1\right]}.
\end{align*}
These equations can be easily integrated to give
\begin{equation}
\tau = t_Q\sqrt{D_1D_2}, \qquad \phi = t_Q D_2,
\end{equation}
where we have also chosen the sign of $\tau$ coinciding with the sign of
$t_Q$. These solutions hold close to the bounce, hence we chose a very
small value of $t_Q$, for instance $t_Q^{\mathrm{ini}} \propto \pm
\mathcal{O}(10^{-50})$, to start the numerical evolution of
Eqs.~\eqref{dalphadt} and \eqref{dphidt}, using the new time $t_Q$. In
order to do that, one must know $D_1$ and $D_2$, hence $\alpha_b, d,
\sigma$ must be given. The above choice of initial $t_Q$ gives well
defined numerical results for the whole range of parameters studied in
this work. The time variable $t_Q$ is then used to solve the quantum
dynamics until the matching with the classical phases, in the contracting
branch and in the expanding branch. For $d<0$, the positive time
direction in the integration moves the solution to the DE branch, while
for $d>0$, the positive time direction moves the solution to the branch
without DE. From there on, we have two possibilities to parametrize the
matching, depending on whether there is a DE behavior in the classical
dynamics or not.

\subsection{Matching with matter domination scale}

Evolving the quantum era as described above, we arrive at a nearly classical
evolution with stiff matter behavior at some $a_\pm = \ln(\alpha_\pm)$ with
$x(a_\pm) = \pm [1 - \epsilon_\pm(a_\pm)]$. At this point, we match the quantum
evolution with the classical one.
In order to control this matching and its cosmological meaning,
we will write $a_\pm$ and its corresponding $\epsilon_\pm (a_\pm)$ in terms of the number of
e-folds between the bounce and the matter-fluid domination, where $x = 1
/ \sqrt{2} + \epsilon_{\lambda}$, with $0 < \epsilon_{\lambda} \ll 1$. We will suppose
that the free constant $\bar{a}_0$ belongs to the infinity open set of real numbers satisfying $x(\bar{a}_0) = 1
/ \sqrt{2} + \epsilon_{\lambda}(\bar{a}_0)$, with $0 < \epsilon_{\lambda}(\bar{a}_0) \ll 1$.

The first step is to impose continuity of the Hubble function at the
matching point. Expanding Eq.~\eqref{C1_sol_cl} about $x_\pm$ gives
\begin{equation}\label{C2Omegad}
\left(\frac{H_\pm}{H_0}\right)^2 \approx
\frac{C_1}{\gamma_\mp^2}\left(\frac{{\bar{a}}_0}{a_\pm}\right)^6.
\end{equation}
Equating this expression to Eq.~\eqref{cl_lim_H_qt} yields
\begin{equation}\label{detC1}
C_1 = \frac{R_{H}^2}{\lP^2} \frac{d^2 \gamma_\mp^2}{\bar{a}_0^6}.
\end{equation}
This equation relates the free constant $C_1$ of the classical system to $\bar{a}_0$.
As the scale factor at the bounce $a_b$ was already chosen, giving the physical parameter
\begin{equation}\label{Xb}
\Xb \equiv \frac{{\bar{a}}_0}{\ab},
\end{equation}
is equivalent to fixing $C_1$. The parameter $\Xb$ yields the number of e-folds from the bounce to the moment 
of the matter-fluid domination determined by $\bar{a}_0$, which, as commented above, is still rather arbitrary:
the only constraint on it is to satisfy $x(\bar{a}_0) = 1/ \sqrt{2} + \epsilon_{\lambda}(\bar{a}_0)$,
with $0 < \epsilon_{\lambda}(\bar{a}_0) \ll 1$. 

For the second constant $C_2$, we obtain from Eq.~\eqref{aapm} that
\begin{equation}\label{detC2}
C_2 = \frac{\gamma_\mp^2}{2^{\gamma_\mp} \epsilon_\pm^{\gamma_{\pm}}} \left(\frac{a_\pm}{{\bar{a}}_0}\right)^3.
\end{equation}
This equation relates the matching point $a_\pm$ and its corresponding $\epsilon_\pm (a_\pm)$ to $C_2$.
Note, however, that the end of the quantum evolution does not designate any specific value of $a_\pm$
as long as the quantum evolution stays very close to the stiff matter classical evolution.
Hence, all points where $0 < \epsilon_\pm \ll 1$ are acceptable.
Here lies the ambiguity of the matching.

We could arbitrarily choose $a_\pm$ in order to fix $C_2$.
However, we will do the reverse: we will connect $C_2$ with sensible cosmological parameters
associated with physical features of the classical branch, and after
a judicious choice of them determining $C_2$, we use Eq.~\eqref{detC2} to finally find the matching point $a_\pm$.
This can be done as follows: close to the matter-fluid epoch, the zero order term of the Hubble
function reads
\begin{equation}
\label{relabel0}
\left(\frac{H}{H_0}\right)^2 \approx \frac{C_1}{C_2\gamma_{-}^{\gamma_+}\gamma_{+}^{\gamma_-}}\left( \frac{{\bar{a}}_0}{a}\right)^3.
\end{equation}
The above result motivates the definition of the arbitrary constant
\begin{equation}\label{relabel}
\Omega_d = \frac{C_1}{C_2 \gamma_{-}^{\gamma_+}\gamma_{+}^{\gamma_-}} = \frac{R_{H}^2}{\lP^2} \frac{d^2 \gamma_\mp^2}{C_2 \gamma_{-}^{\gamma_+}\gamma_{+}^{\gamma_-}\bar{a}_0^6}.
\end{equation}
This constant is very useful, as long as it gives a precise meaning to $\Xb$. Indeed, from Eq.~\eqref{relabel0}
and Eq.~\eqref{relabel}, we get
$$
H^2(a = {\bar{a}}_0) \approx H_0^2\Omega_d.
$$
The parameter $\Xb$ can now be understood as yielding the number of
e-folds between the bounce and the moment where the Hubble radius is $R_H
/ \sqrt{\Omega_d}$. Hence, once $\Omega_d$ is given, $\Xb$ acquires a very precise meaning.

In terms of the physical variables $\Omega_d$ and $\Xb$, the constant $C_2$ reads,
\begin{equation}
C_2 = \frac{R_{H}^2}{\lP^2} \frac{d^2 \gamma_\mp^2}{\Omega_d \gamma_{-}^{\gamma_+}\gamma_{+}^{\gamma_-}\ab^6\Xb^6}.
\end{equation}
This expression is completely determined by our choices of quantum
initial condition $a_b$ and the constants $\Omega_d$ and $\Xb$. Plugging
it into Eq.~\eqref{detC2}, we obtain our matching time
\begin{equation}\label{matcha}
\frac{a_\pm}{\epsilon_\pm^{\gamma_\pm/3}} \approx \frac{1}{\Xb \ab}\left(\frac{R_{H}^2d^2  2^{\gamma_\mp}}{\lP^2\gamma_{-}^{\gamma_+}\gamma_{+}^{\gamma_-} \Omega_d}\right)^{1/3}.
\end{equation}
Then, given a value for $\Xb$ and $\Omega_d$, with the cosmological
meanings described above, we must evolve the quantum branch until some
values of $\alpha_\pm = \ln a_\pm$ and $\epsilon_\pm$ where
Eq.~\eqref{matcha} is satisfied. From this point on, we continue using
the classical equations \eqref{sist1c} with these matching values for
$\alpha_\pm$ and $x_\pm$ as initial conditions.

Summarizing, from the classical dynamics we introduced four (redundant)
constants to control the initial conditions $({\bar{a}}_0,\; H_0,\;
C_1,\; C_2)$. We chose $H_0$ to match today's value of the Hubble
parameter in order to have the problem scaled to the perturbation scales
of cosmological interest. To fix the other arbitrary parameters, we
studied the classical solution around the matter-fluid critical point,
introducing the matter parameter $\Omega_d$, which only serves to give a
meaning to $\Xb$. Note that the matching~\eqref{matcha} depends only on
the product $\Xb^3\Omega_d$, evincing that one of these parameters is
arbitrary. The first actual initial condition we impose by matching the
value of the Hubble function in the end of a quantum branch with its
value in the beginning of a classical branch. The second initial
condition is, however, not completely determined by the quantum phase, as
we can choose any small value of $\epsilon_\pm$ as we want. For this
reason, we choose a value for $\Xb$ (keeping $\Omega_d$ fixed), which has
a clear cosmological meaning, to
completely determine $a_\pm$ and $\epsilon_\pm$ through Eq.~\eqref{matcha},
and the subsequent system evolution .

Finally, it is useful to study the allowed values for $\Xb$. Using for the moment $\Omega_d = 1$, i.e., choosing $\Xb$ to represent the number of e-folds between the bounce and the instant where the Hubble radius matches today's value, we get
\begin{equation}\label{matcha0}
\Xb \approx (a_b a_\pm)^{-1}\left(\frac{R_{H}^2d^2  2^{\gamma_\mp}\epsilon_\pm^{\gamma_{\pm}}}{\lP^2\gamma_{-}^{\gamma_+}\gamma_{+}^{\gamma_-}}\right)^{1/3}.
\end{equation}
The value of $\Xb$ is inversely proportional to $a_ba_\pm$, thus, to maximize $\Xb$ we need to minimize $a_ba_\pm$. Using the fact that $\alpha_b > 0$, we get $a_b = 1$ as the minimum value of $a_b$. Assuming that the quantum phase is fast enough so that it
has approximately zero e-folds of duration, we get $a_\pm = 1$ as the minimum value of $a_\pm$.
Then, ignoring the order one factors we get
\begin{equation}
\Xb \lesssim 10^{40} d^{2/3} \epsilon_\pm^{\gamma_{\pm}/3},
\end{equation}
which shows, that given the value of $d$, we have a maximum for the number
of e-folds. This fact is very relevant for the perturbations,
since their amplitudes are determined partially by $\Xb$. Note also that
this kind of constraint is present in any matter bounce, and it can make
difficult to generate enough amplitudes for the perturbations without
approaching too much the Planck scale. Moreover, $\Xb$ is proportional to
$\epsilon_\pm^{\gamma_{\pm}}$, thus, a long quantum branch
($\epsilon_\pm^{\gamma_{\pm}}$ very small) results in a shallow bounce.

\subsection{Matching with the dark energy scale}

For trajectories containing the DE epoch, we have an alternative way to
give meaning for a reference scale. We can choose ${\bar{a}}_0$ to
represent the exact point where $w = -1$, i.e., $x = 0$. At this point we
have
\begin{equation}\label{matchde}
\left[\frac{H(a={\bar{a}}_0)}{H_0}\right]^2 = 2C_1 \equiv \Omega_\Lambda, \qquad 2C_2 = 1,
\end{equation}
where we have introduced the parameter $\Omega_{\Lambda}$ (in the same
way and with similar characteristics to $\Omega_d$). There is an
important distinction to make in comparison with the other case. Here, we
are matching with a fixed point in time, whereas, in the last
matter-fluid we can match to any time where $0 < \epsilon_\lambda \ll 1$.
For this reason, the value of $\Omega_{\Lambda}$ completely determines
the matching and in this case $\Xb$ is obtained from it. For
$\Omega_\Lambda = 1$, DE domination takes place around our present Hubble
time.

Since we have fully determined $C_1$ and $C_2$ all we need to do is
substitute them on Eqs.~\eqref{detC1} and \eqref{detC2} from which we obtain the matching condition
\begin{equation}\label{matchde2}
\frac{a_-}{\epsilon_-^{\gamma_-/3}} \approx \left(\frac{R_H \vert
d\vert 2^{\gamma_+}}{\lP\sqrt{2\Omega_{\Lambda}}\gamma_+}
\right)^{1/3},
\end{equation}
where we specialized in the ${}_-$ branch since it is the only one
containing a DE phase. The number of e-folds in this case is given by the
logarithm of
\begin{equation}
\Xb \approx \frac{1}{\ab}\left[\frac{2 R_H^2 \gamma_{+}^2 d^2 }{\lP^2\Omega_{\Lambda}} \right]^{{1}/{6}}. \label{a_0_DE}
\end{equation}
It is worth noting that the number of e-folds between the bounce and the
$\vert H\vert = H_0$ scale, assuming $\Omega_\Lambda = 1$, is different for the different matching
procedures, showing the asymmetry of the model. Note also that the main
factor in determining $\Xb$ in the matter-fluid matching is
$(R_H/\lP)^{2/3}$ while for the DE matching we have $(R_H/\lP)^{1/3}$,
i.e., the DE matching produces a smaller number of e-folds in the branch
where it is applied. It is also clear from the equation above that larger
(smaller) $\Omega_\Lambda$ results in fewer (more) e-folds between the DE
phase and the bounce.

One last comment: when the DE phase happens during expansion (the case
\B) it is natural to use this procedure to match, using $\Omega_{\Lambda}
\approx 1$, which guarantees us that we could model the current
accelerated expansion using this field. However, if the DE phase happens
during the contraction (case \A), we do not have any reason to choose
\emph{a priori} a given scale to match.

\subsection{Summarizing the background reconstruction}

As explained in this section, the numerical integration used to construct
the background model is initiated at the bounce itself, using the quantum
guidance equations. For that, one should give the values of $\alpha_b$,
$d$ and $\sigma$. The system is evolved until reaching the classical
limit, where it is matched with the classical evolution. This matching is
controlled by the cosmological parameters $\Omega_\Lambda$ and $\Xb$. In
the branch with a DE phase, the quantum evolution is halted when
Eq.~\eqref{matchde2} is satisfied, while in the branch without a DE phase
the quantum dynamics is stopped when condition~\eqref{matcha} is reached.
This gives the values of $\alpha_\pm$ and $x_\pm$ to be used as initial
conditions to the subsequent integration of the classical equations
\eqref{sist1c}.

Hence, the complete collection of parameters needed to fix the background
model is ($\alpha_b$, $d$, $\sigma$, $\Omega_\Lambda$, $\Xb$), all of
them with clear cosmological significance. For $d>0$ we have case \A, and
the DE phase is in the contracting phase, while for $d<0$ we are in case
\B, and the DE phase is in the expanding phase.

As a last remark, the classical branch could, in principle, be calculated
using Eq.~\eqref{sist1c}, and all other quantities are obtained by simple
integrals of $x$. However, the variable $x$ is not well behaved
numerically, for instance, to represent a point very close to $S_+$, say
$x = 1 - 10^{20}$, we would need at least a $20$ decimal digits floating
point number, well beyond the default double precision (about $16$
decimal digits), present in today's computers. To avoid this numerical
pitfall, we make the following change of variables
\begin{equation}
x_\pm(q_\pm) = \pm \left(1-\frac{\gamma_\mp q_\pm}{1+q_\pm}\right),
\end{equation}
which maps the $x$ range $(\pm1, 1/\sqrt{2})$ into the numerically well
defined interval $(0,\; \infty)$.\footnote{The default double precision
float point number can represent numbers from $\approx 10^{-300}$ to
$\approx 10^{+300}$.} For this reason we solve numerically
Eq.~\eqref{sist1c} written in terms of $q_\pm$ for the paths
ending/beginning on $S_\pm$.

\section{Numerical solutions for the background}
\label{sec_num_back}

\begin{figure}[ht]
\begin{center}
\includegraphics[width=0.5\textwidth]{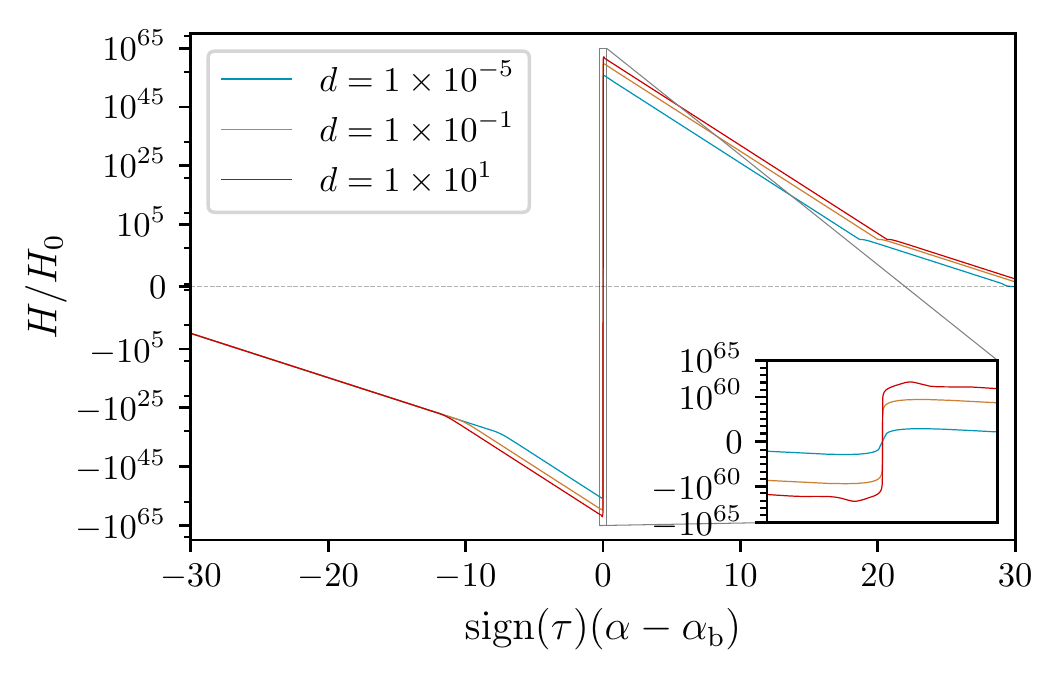}
\end{center}
\caption{For case \B, the dependence of the background dynamic with the
parameter $d$. Smaller values of $d$ implies in longer matter duration.
This behavior is also noticed in case \A. We included a zoom-in of the
narrow interval around the bounce, showing the behaviors of $H$ using
different values of $d$.} \label{Fig:bckg_d}
\end{figure}

In this section we will explore the parameter space of the theory, and
see its influence in the complete background behavior, which will be used
in the perturbative analysis. When one parameter is varied in the figures
below, the bold values in Tab.~\ref{tab_par} designate the other
parameters which get fixed in the corresponding figures. For example, the 
different curves
appearing in Fig.~\ref{Fig:bckg_d} for the case \B corresponding to the
different values of $d$ present in Tab.~\ref{tab_par} were calculated
with the parameters $\sigma = 0.5$, $\alpha_{\mathrm{b}} = 10^{-40}$,
${\mathcal{X}}_{\mathrm{b}}= 10^{30}$ and $\Omega_{\Lambda} = 1$. Also, since we
are interested in perturbations at scales near the Hubble radius today,
we from here on use $\Omega_d = 1$. 

\begin{table}
\begin{tabular}{|c|c|c|c|c|}
\hline
$d$ & $\sigma$ & $\alphab$ & $\Xb$ &
$\Omega_{\Lambda}$ \\
\hline
&&&&\\[-0.8em]
$10^{-5}$ & $5 \times 10^{-2}$ & $\mathbf{10^{-40}}$ & $10^{20}$ & $\mathbf{1}$
 \\
\hline
&&&&\\[-0.8em]
$\mathbf{10^{-1}}$ & $\mathbf{5 \times 10^{-1}}$ & $10^{-5}$ & $10^{25}$ &
$10^{20}$ \\
\hline
&&&&\\[-0.8em]
$10$ & $5$ & $1$ & $\mathbf{10^{30} }$ & $10^{40}$ \\
\hline
\end{tabular}
\caption{The parameters of the numerical solutions,
Figs.~\ref{Fig:bckg_d} to \ref{Fig:bckg_Xb}. The bold values in the table
are fixed when one parameter is varied. In all examples we considered
$\Omega_d = 0.3$} \label{tab_par}
\end{table}

\begin{figure}[ht]
\begin{center}
\includegraphics[width=0.5\textwidth]{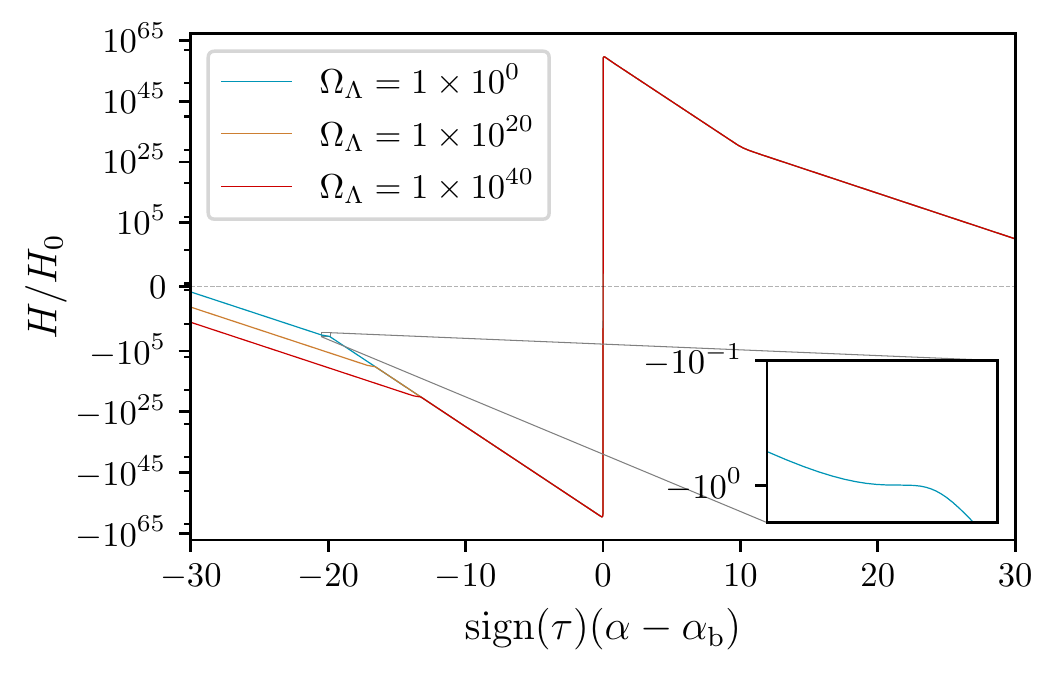}
\end{center}
\caption{For case \A, the dependence of the DE epoch with the parameter
$\Omega_{\Lambda}$. Smaller values lead to earlier DE epochs. This
behavior is also noticed in case \B, but, since in case \B the DE epoch
happens in the expanding phase, smaller values of $\Omega_{\Lambda}$ will
imply a later DE epoch. We included a zoom-in of the narrow interval
around the DE phase, showing the constance of $H$ around it.}
\label{Fig:bckg_Omega_L}
\end{figure}

\begin{figure}[ht]
\begin{center}
\includegraphics[width=0.5\textwidth]{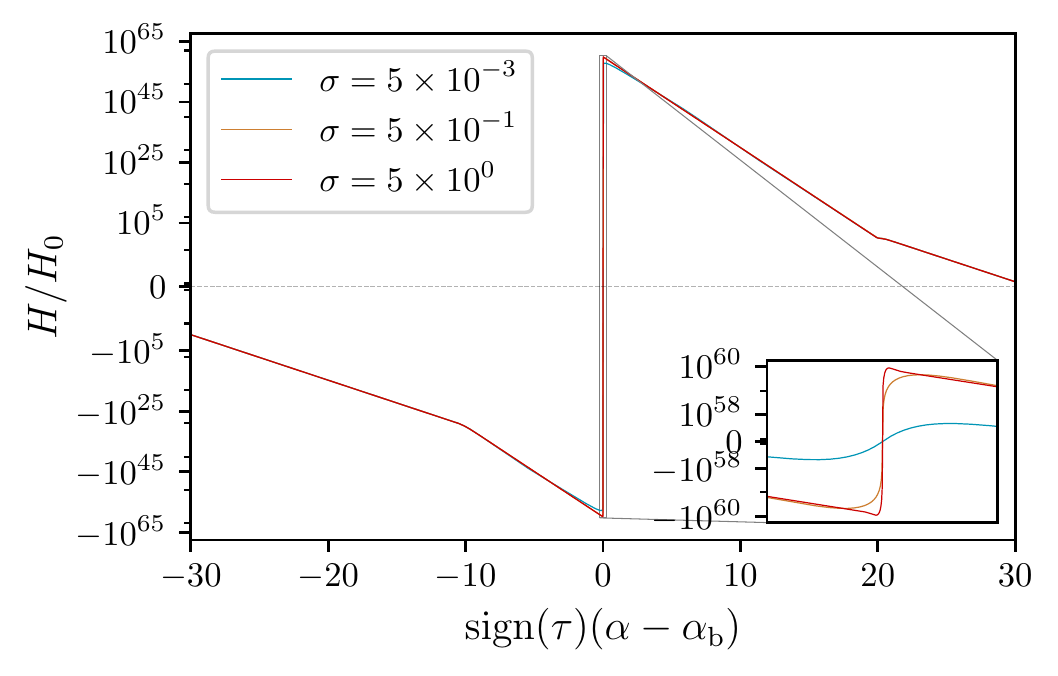}
\end{center}
\caption{For case \B, the dependence of the $H_{\mathrm{max}}$ with
$\sigma$. $H_{\mathrm{max}}$ is an important parameter in order to
determine the validity of the canonical quantization, since we should
maintain the energy scale of the bounce below Planck scale.}
\label{Fig:bckg_sigma}
\end{figure}

\begin{figure}[ht]
\includegraphics[width=0.5\textwidth]{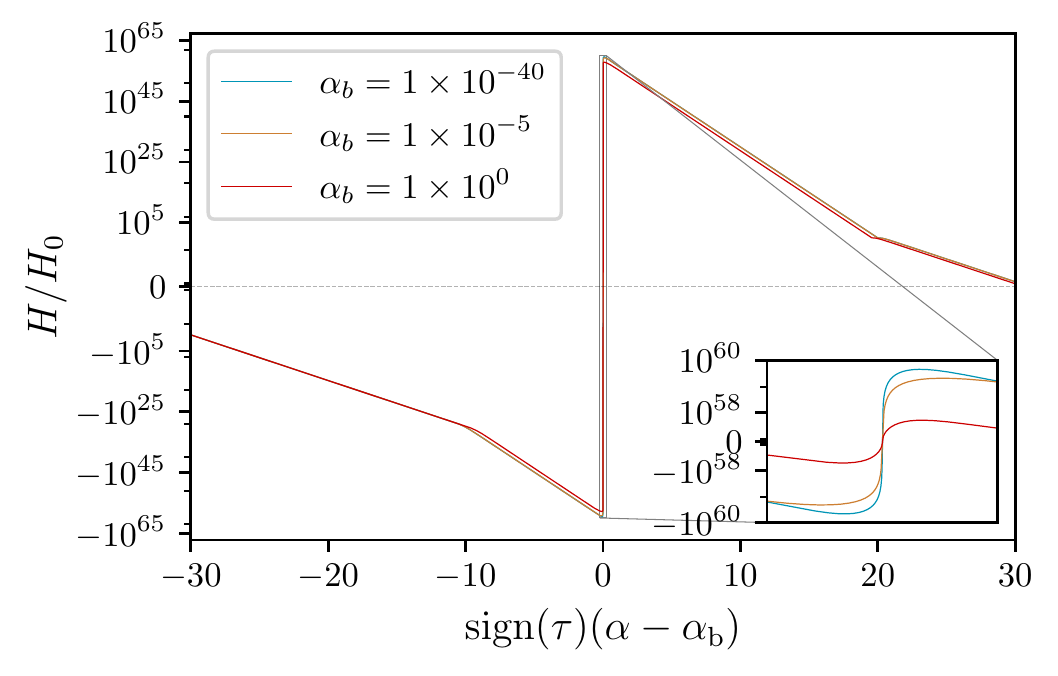}
\caption{For case \B, the dependency of the bounce energy scale with the
minimum scale factor. Smaller values of $\alphab$ allows a
longer contraction that increases $\vert H_{\mathrm{max}}\vert$. }
\label{Fig:bckg_alphab}
\end{figure}

\begin{figure}[ht]
\includegraphics[width=0.5\textwidth]{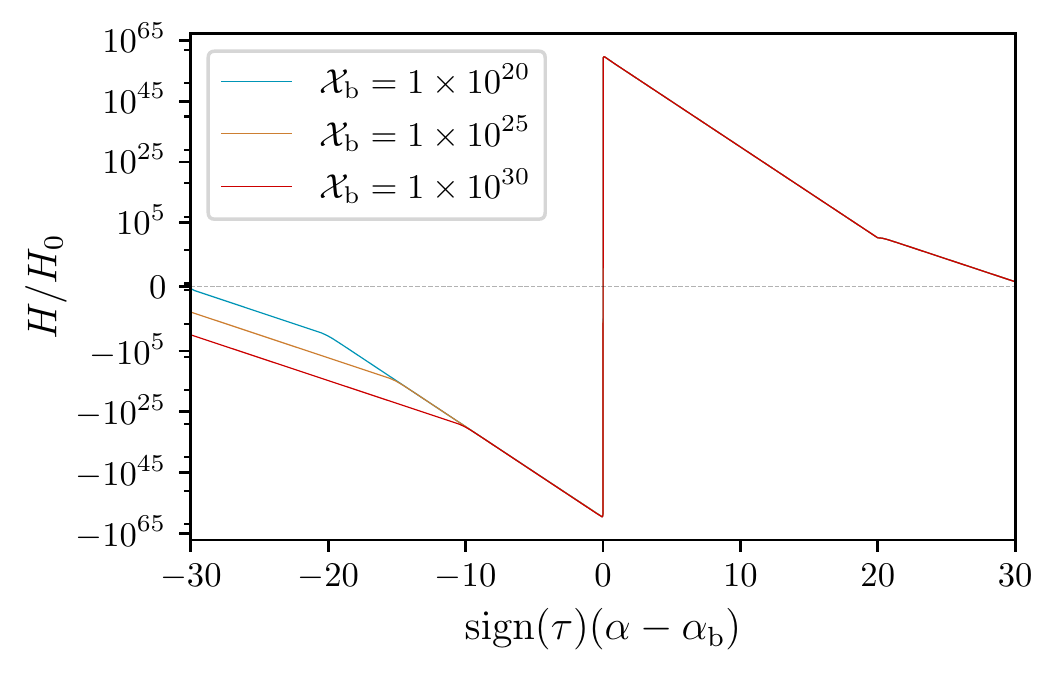}
\caption{For case \B, the duration of matter epoch is longer for bigger
values of $\Xb$. } \label{Fig:bckg_Xb}
\end{figure}

In what concerns the study of the cosmological perturbations, the
background dynamics can be fully understood by the plot of $H$ with
$\alpha$, Figs.~\ref{Fig:bckg_d} to \ref{Fig:bckg_alphab}. We show
clearly the bounce asymmetry by choosing the horizontal axis as being
$\mathrm{sign}(\tau)(\alpha-\alphab)$. In that way, the negative interval
represents the contracting phase while the positive interval represents
the expanding phase. For a perfect fluid with $p = w \rho$, the evolution
of $H$ is
\begin{equation}
 \ln \vert H\vert \propto \frac{3}{2}(w + 1) \alpha,
\end{equation}
thus, in the intervals with a well defined $w$ (stiff and matter-fluid
phases) $H$ behaves as a power-law. In the matter epoch, the effective
EoS of the scalar field is $w = 0$, and in the stiff matter one it is
$w=1$, implying in different slopes in Figs.~\ref{Fig:bckg_Omega_L} to
\ref{Fig:bckg_Xb}. The duration of the epochs are connected with the size
of the Universe at which we see the transition from one slope to another.
The closer to the bounce that transition happens, longer is the matter
epoch. This is very important since we are interested in controlling the
matter contraction to confirm its influence in the relevant mode scales.

The DE epoch happens when $x = 0$ corresponding to a short plateau
between the matter and stiff-matter phases, for example, in
Fig.~\ref{Fig:bckg_d} around $\alpha = 20$.\footnote{This could be
deduced directly from Eq.~\eqref{a_0_DE} since $R_H/\lP \approx 10^{60}$
and all other constants are of order one.} In Fig.~\ref{Fig:bckg_Omega_L}
we included a zoom-in for this interval, showing the transient DE phase.

We plotted Figs.~\ref{Fig:bckg_d} to \ref{Fig:bckg_Xb} for case \B,
however, equivalent figures for case \A can be obtained by simple mirror
symmetry
$$
\mathrm{sign}(\tau)(\alpha-\alphab) \rightarrow - \mathrm{sign}(\tau)(\alpha-\alphab)
$$
As we have mentioned before, it makes no sense to change
$\Omega_{\Lambda}$ away from $\approx 1$ in case \B, since it is an
observational constraint and we expect to use this transient DE phase to
model the current accelerated expansion. On the other hand, in case \A,
this is exactly the parameter we are interested in order to
study
perturbations in a DE epoch. Therefore, Fig.~\ref{Fig:bckg_Omega_L} is a
plot from case \A.

The bounce happens in $\alpha = \alphab$, and the two peaks are the
highest values of $H$ reached by the system, further on referred as
$H_{\mathrm{max}}$.  The peaks happen when $\dot{H} = 0$ and we can use
them to define the duration of the bounce, $\delta_{\mathrm{b}}$. They
can be better noticed in the zooms depicted in Figs.~\ref{Fig:bckg_d} and
\ref{Fig:bckg_sigma}. The closer the peaks are in the plots, the smaller
is $\delta_{\mathrm{b}}$ (faster bounce).

In what concerns the quantum solutions, the variation of the parameters
$d$, $\sigma$ and $\alpha_{\mathrm{b}}$ directly changes the time and
energy scales of the bounce. When increasing $d$, the frequency in
Eqs.~\eqref{dalphadt} and \eqref{dphidt} is higher and it is possible
that the background oscillates close to the bounce,
Fig.~\ref{Fig:bckg_d}.

Another important influence of $d$ is in the duration of the matter phase
in contraction. Using Eq.~\eqref{aalambda}, we can mark the onset of the
matter-fluid phase by choosing a fixed $\epsilon_\lambda$ (for example
$10^{-4}$). Then, it is clear from Eq.~\eqref{C2Omegad} that $a_\lambda$
(marking the beginning of the matter-fluid phase) is proportional to
$\vert d\vert^{2/3}$. Hence, larger $d$ yields longer stiff-matter
phases. This effect can be seen in Fig.~\ref{Fig:bckg_d}. In the
expanding era, which contains a DE phase, the same Eq.~\eqref{aalambda}
can be used, but now $C_2 = 1/2$, and the connection to $d$ is through
${\bar{a}}_0$ given by Eq.~\eqref{a_0_DE}. Consequently, $a_\lambda
\propto \vert d\vert^{1/3}$. In the expanding era of
Fig.~\ref{Fig:bckg_d} we can see that the DE plateau shifts by a smaller
$\delta\alpha$ than the matter-fluid era shift in the contraction phase.

The parameter $\sigma$ is relevant only in the quantum phase.
Figure~\ref{Fig:bckg_sigma} shows that larger $\sigma$'s implies in
higher energy and in shorter time scales in the bounce. This can be
easily understood looking again to Eqs.~\eqref{dalphadt} and
\eqref{dphidt}. The hyperbolic functions have the argument
$\sigma^2\alpha \phi$ [see for instance Eq.~\eqref{cl_lim_x_qt}] and they
saturate when the argument is of the order of $11$ [$x \approx \coth(11)
\approx 1 + \mathcal{O}{\left(10^{-10}\right)}$]. Hence, a larger value
for $\sigma$ leads to a faster saturation of the hyperbolic functions
and, consequently, to an earlier stiff-fluid epoch. Furthermore, in the
small $\alpha$ and $\phi$ approximation, the value of
$|H_{\mathrm{max}}|$ is also proportional to $\sigma$, meaning that
larger values of $\sigma$ generate more energetic bounces.

A similar argument leading to higher values of $\vert
H_{\mathrm{max}}\vert$ can be used when analyzing the influence of
$\alphab$, Fig.~\eqref{Fig:bckg_alphab}. More profound bounces imply in a
longer stiff-matter epoch, as we can see in the transitions of the slopes
in Fig.~\eqref{Fig:bckg_alphab}. During that period $\vert H\vert$ has
more time to increase leading to high-energy scale bounces.

Finally, in Fig.~\eqref{Fig:bckg_Xb} we observe that $\Xb$, controlling
the matching point between the classical and quantum branches, influences
the duration of the dust domination era. This turns out to be crucial for
the perturbations, not only because the power-spectrum slope depends on
which fluid dominates when the mode leaves the Hubble radius, but also
because the spectrum amplitude depends on how long the mode evolves in
each fluid-like epoch.

\section{Primordial perturbation}
\label{sec_pert}

The perturbed Einstein's equations  can be recast in a very simple and
objective manner combining the scalar perturbation in the metric and in
the matter component by means of the gauge invariant curvature
perturbation $\zeta$. The action that gives the dynamic for the
perturbations comes from the first-order perturbed Einstein-Hilbert
action and reads~\cite[Eq.~(66)]{Falciano2013}
\begin{equation}
S = \int\dd\tau\, \dd x^3 \frac{z^2}{2}\left({\zeta}^{\prime2} + \frac{N^2\zeta \Delta \zeta}{a^2}\right),\;\; z^2 \equiv \frac{3a^3x^2}{\kappa^2N},
\label{action_v}
\end{equation}
where $\Delta$ is the spacial Laplacian, $\zeta$ is the gauge invariant
dimensionless curvature perturbation, and the time operator is defined by
$\prime \equiv \dd/\dd\tau$.\footnote{Note that, our definition of $z$ is
different from the one in Ref.~\cite[Eq.~(66)]{Falciano2013}, $z^2 =
1/(N\kappa^2{\bar{z}}^2)$, where ${\bar{z}}$ denotes the $z$ appearing in
Ref.~\cite[Eq.~(66)]{Falciano2013}.} The equation of motion for $\zeta$
is obtained by the variational principle, and after the field
decomposition it reads
\begin{equation}
\zeta^{\prime\prime}_k + 2\frac{z^\prime}{z}{\zeta_k^\prime} + \frac{N^2k^2}{R_H^2a^2}\zeta_k = 0.
\label{EOM_v}
\end{equation}
In the above equations, we used the dimensionless time $\tau$
[Eq.~\eqref{del_tau}],
$$ N = \frac{\dd t}{\dd\tau} = \frac{\tau}{H}.$$
Here $k$ is measured in units of $R_H^{-1}$,\footnote{The eigenvalues of
the Laplacian $\Delta$ are given by $-R_H^{-2}k^2$.} $z$ and $\zeta_k$
have dimensions of $\mathrm{lenght}^{-3/2}$ and $\mathrm{lenght}^{3/2}$
respectively.

We are in the domain of linear perturbation theory, in which the scalar,
vector and tensor perturbations decouple. The tensor perturbation
$h_{ij}$, whose amplitude of any of its two polarizations will be
refereed by $h$, presents a similar action, i.e.,
\begin{equation}
S = \int\dd\tau\, \dd x^3 \frac{z_h^2}{2}\left(h^{\prime2} + \frac{N^2 h \Delta h}{a^2}\right),\;\; z_h^2 \equiv \frac{a^3}{4\kappa^2N}.
\label{action_h}
\end{equation}
Its equation of motion can be easily deduced from the action above,
\begin{equation}
h_k^{\prime\prime} + 2\frac{z_h^\prime}{z_h}{h_k^\prime} + \frac{N^2k^2}{R_H^2a^2}h_k = 0,
\label{eom_tensor}
\end{equation}
where $h_k$ also has dimension of $\mathrm{length}^{3/2}$. The above
formulation can be found in the literature, for instance
in~\cite{Mukhanov1992, Peter2009, Brandenberger2003}. Here we introduced
only the necessary information in order to define the observational
probes that we will calculate. The quantities constrained by the
observations are the power spectra,
\begin{align}
\Delta_{\zeta_k} &\equiv \frac{k^3 \vert\zeta_k\vert^2}{R_H^3 2 \pi^2} = \frac{\lP^2}{R_H^2}\frac{4k^3 \vert\tzeta_k\vert^2}{3\pi},  \\
\Delta_{h_k} &\equiv \frac{k^3\vert h_k\vert^2}{R_H^3 2\pi^2} = \frac{\lP^2}{R_H^2} \frac{16k^3\vert \thg_k\vert^2}{\pi},
\end{align}
where we introduced the dimensionless mode functions
\begin{equation}\label{dimless}
\zeta_k \equiv \sqrt{\frac{\kappa^2R_H}{3}}\tzeta_k,\qquad h_k \equiv \sqrt{4\kappa^2 R_H}\thg_k.
\end{equation}
The spectral indexes for the scalar curvature perturbation $\zeta$ and
tensor perturbation $h$ labeled by the subscripts s and T are,
respectively,
\begin{equation}
 n_{\mathrm{s, T}} - 1 \equiv  \left.\frac{\dd \log (\Delta_{\zeta_k, h_k})}{\dd\log k}\right\vert_{k = k_*}
\end{equation}
and the  tensor-to-scalar ratio
\begin{equation}
r \equiv 2 \left.\frac{\Delta_{h_k}}{\Delta_{\zeta}}\right\vert_{k = k_*},
\end{equation}
where the factor $2$ comes to account for the two polarizations of the
tensor perturbation. We use the same pivotal scale as
in~\cite{PlanckCollaboration2015b}, $k_* = 0.05R_H\;\mathrm{Mpc}^{-1}$.
The latest Planck release estimates for long wave-lengths
$\Delta_{\zeta_k} \approx 10^{-10}$, $n_{\mathrm{s}} \approx 0.96$ and $r < 0.1$~\cite{PlanckCollaboration2015b}.

Following the results from \cite{Peter2007, Peter2008}, the spectral
index for the modes entering the horizon during the domination of a fluid
with EoS $w$ is
\begin{equation}
n_{\mathrm{s}} = 1 + \frac{12w}{1 + 3 w}.
\end{equation}
In case \B, the spectrum is scale invariant when entering the horizon
while $w = 0$, matter epoch. Depending on the duration of the matter
contraction, some very small scale modes may enter during the transition
to the stiff-matter phase, $w = 1$, leading to a blue spectrum. For case
\A, besides the above two possibilities, there is also the influence of
the transient DE epoch in large scale modes. Hence, although case \A is
more academic, it should be quite interesting to evaluate the exact
spectral index in this case in order to estimate the impact of the
presence a transient DE in the contracting phase of a matter bounce in
the standard results.

\subsection{Initial vacuum perturbations}

It is usually proposed in current cosmological models that the
inhomogeneities in the Universe have their origin in primordial vacuum
quantum fluctuations. In the inflationary scenario, the exponential
growth of the scale factor are responsible for amplifying those quantum
fluctuations. After a $60$ e-fold expansion they have enough amplitude to
fit the CMB observations.

Bouncing models assume the same mechanism for the origin of
inhomogeneities, but replaced in the far past of the contracting phase.
Some scenarios may find difficulty in providing the Minkowsky vacuum as
initial conditions. This is the case when the cosmological constant is
considered \cite{Maier2011}.

In the present case, the scalar field behaves like a matter fluid and the
usual quantization of the adiabatic vacuum fluctuations in a Minkowsky
space-time coincides with the WKB solution with positive energy for the
Mukhanov-Sasaki variable $v \equiv z\zeta$. The equation of motion
\eqref{EOM_v} can be written
\begin{equation}
v_k^{\prime\prime} + w_k^2 v_k = 0,
\label{eom_v}
\end{equation}
where,
\begin{equation}
\label{def_w_k}
w_k^2(\eta, k) \equiv \frac{N^2k^2}{a^2R_H^2} - \frac{z^{\prime\prime}}{z}.
\end{equation}
Note that the expressions above reduce to the usual conformal time
equations for $N = a$.

A solution of the above equation can be expressed in terms of the WKB
approximation (see for example~\cite{Peter2009}), which has a certain
limit of validity. Let us define,
\begin{equation}
Q_{\mathrm{WKB}} = \frac{3}{4}\left(\frac{w_k^{\prime}}{w_k}\right)^2-
\frac{1}{2}\frac{w_k^ {\prime \prime}}{w_k}.
\end{equation}
In the regime in which
\begin{equation}
\label{cond_WKB}
\left\vert \frac{Q_{\mathrm{WKB}}}{w_k^2}\right\vert \ll 1,
\end{equation}
the solution is
\begin{equation}
\label{sol_wkb}
\tilde{v}_k^{\mathrm{WKB}} \approx \frac{1}{\sqrt{2 w_k}} \e^{\pm i
\int\dd\tau w_k}.
\end{equation}
The matter contraction satisfies  \eqref{cond_WKB} and
Eq.~\eqref{sol_wkb} not only gives the initial conditions but also a good
approximation for the solution of Eq.~\eqref {eom_v} while condition
\eqref{cond_WKB} is satisfied.

For $N^2k^2/(a^2R_H^2) \gg \vert z^{\prime\prime}/z\vert$ the initial
vacuum conditions are reduced to
\begin{align}
v_{\mathrm{ini}} &= \frac{1}{\sqrt{2k}}\sqrt{\frac{aR_H}{N}}, \label{ini_vacuo_1}\\
\left.\frac{\dd{}v}{\dd\tau}\right\vert_\mathrm{ini} &= i \sqrt{2 k}\sqrt{\frac{N}{aR_H}}, \label{ini_vacuo_2}
\end{align}
where we have set the initial phase factor equal to zero. Again, we can
recover the usual vacuum conditions choosing the conformal time lapse
function $N = a$. The tensor modes $h$ can be expressed in terms of the
variable
\begin{equation}
\mu = z_h h,
\end{equation}
which satisfies similar equations as $v$, but with $z_h$. The same
treatment given to the quantization of $v$ can be performed for $\mu$ and
the initial conditions are equivalent for the tensor modes.

In the adiabatic limit, the perturbations are in a high oscillatory
regime and the numerical calculations become contrived. A very common
approach to numerically solve the perturbations is to consider the WKB
solution until $N^2k^2/(a^2R_H^2) > \vert z^{\prime\prime}/z\vert$ and to
switch to the numerical calculation just before the saturation of the
inequality. The problem with this approach is that, by construction, the
WKB approximation is worse and worse as we approximate the saturation
point. Thus, we must balance two problems, first if we start the
numerical evolution very close to the saturation we would need a high
order WKB approximation to get a reasonable initial condition. The high
order WKB approximation needs high order time derivatives of the
background functions, which in turn is badly defined numerically or
involves complicated background functions also numerically error prone.
If we decide to start away from the saturation point we need to deal with
the highly oscillatory period, in which the processing time is longer and
the numerical errors accumulate with the oscillations.\footnote{One
should also note that, the usual approximation, $k^2 = V$, used to
calculate the power-spectrum, underestimates its
amplitude~\cite{Martin2002a}.}  To circumvent the above mentioned
problems, we will use the action angle variables to rewrite the
perturbations dynamics and numerically solve the new sets of equations.
These variables are better suited to the high oscillatory regime and
allows us to use initial conditions away from the saturation point.

\subsection{Action angle variables}

The Action Angle (AA) variables can be used to solve high oscillatory
differential equations~\cite{Vitenti2015, Peter2016, Celani2017}. General
linear oscillatory systems have a quadratic Hamiltonian in the form
\begin{equation}
\mathcal{H} = \frac{\Pi_{\tzeta_k}^2 }{2 m} +  \frac{m \nu^2}{2}\tzeta_k^2,
\label{quad_hamilt}
\end{equation}
where $m$ is the associated ``mass'' of the system and $\nu$ the
frequency. The generalized variable and associated momenta are $\tzeta_k$
and $\Pi_{\tzeta_k}$, respectively.

From action~\eqref{action_v} and the definition of the dimensionless
variables~\eqref{dimless}, we can easily deduce that
\begin{equation}\label{masszeta}
m = \frac{\kappa^2R_Hz^2}{3} = \frac{a^3x^2R_H}{N},\qquad \nu = \frac{N k}{a R_H}.
\end{equation}
The Hamilton equations are
\begin{equation}
\zeta^{\prime}_k = \frac{\Pi_{\zeta_k}}{m}, \qquad \Pi^{\prime} = - m \nu^2 \zeta_k.
\end{equation}
AA variables are based on the adiabatic invariant of oscillatory systems
\cite{LandauLifishits}. A real solution for the Hamilton equations above
can be rewritten in terms of the variables $(I,\;\theta)$, implicit defined
by
\begin{align}
\tzeta^{\mathrm{a}}_k &= \sqrt{\frac{2 I}{m \nu}} \sin \left( \theta \right), \\
\Pi^{\mathrm{a}}_{\tzeta_k} &= \sqrt{2 I m \nu} \cos \left( \theta \right).
\end{align}
Deriving the above expressions and using the Hamilton equations, we find
the equation of motion in terms of the new variables $\theta$ and $I$ are,
respectively,
\begin{align}
 I^{\prime} &= - I \frac{(m \nu)^{\prime}}{m \nu} \cos \left( 2 \theta \right),
 \label{eom_I}\\
\theta^{\prime} &= \nu + \frac{1}{2} \frac{(m \nu)^{\prime}}{m \nu}  \sin \left(2
\theta\right). \label{eom_theta}
\end{align}
The second real solution $\zeta^\mathrm{b}$ introduces another pair of AA
variables,  $(J,\;\psi)$, again implicit defined by
\begin{align}
\tzeta^{\mathrm{b}}_k &= \sqrt{\frac{2 J}{m \nu}} \sin \left( \psi \right),
\\
\Pi_{\tzeta_k}^{\mathrm{b}} &= \sqrt{2 J m \nu} \cos \left( \psi \right).
\end{align}
They follow the same equations of motion
\begin{align}
 J^{\prime} &= - J \frac{(m \nu)^{\prime}}{m \nu} \cos \left( 2 \psi \right)
\label{eom_J},\\
\psi^{\prime} &= \nu + \frac{1}{2} \frac{(m \nu)^{\prime}}{m \nu}  \sin \left(2
\psi \right). \label{eom_psi}
\end{align}
The final complex solution is defined by
$$\tzeta_k = \frac{\tzeta_k^{\mathrm{a}} +i\tzeta_k^{\mathrm{b}} }{2i},
$$
with the real and imaginary parts satisfying the normalization condition
imposed by the initial quantum vacuum perturbations, i.e.,
\begin{equation}
\sqrt{I J} \sin \left( \psi - \theta \right) = 1
\label{norm_cond}
\end{equation}
We will define the variables $\epsilon$, $\bar{\theta}$ and $\Delta
\theta$ in order to re-write the above equations so the constraint will
be automatically satisfied all along the evolution. Defining
\begin{align}
\sinh\left( \epsilon\right) &= \cot \left( \Delta \theta\right),
\label{def_e_pt2}\\
\Delta \theta &= \psi - \theta, \label{def_delta_teta}\\
\bar{\theta} &= \frac{\psi + \theta}{2}, \label{def_teta_barra}
\end{align}
we can easily demonstrate the relations
\begin{align}
 \sqrt{I J} &= \cosh \left( \epsilon \right), \label{def_e_pt1}  \\
\sin \left( \Delta \theta \right) &= \frac{1}{\cosh \left( \epsilon \right)}, \label{eq1} \\
\cos \left( \Delta \theta \right) &= \tanh \left( \epsilon \right). \label{eq2}
\end{align}
Differentiating Eqs.~\eqref{def_teta_barra} and \eqref{def_e_pt1} and
using Eqs.~\eqref{eq1} and  \eqref{eq2} to rewrite $\psi$ and $\theta$ in
terms of $\epsilon$ and $\bar{\theta}$ we find
\begin{align}
\bar{\theta}^{\prime} &= \nu + \frac{(m \nu)^{\prime}}{m \nu} \tanh \left(
\epsilon \right) \sin\left(\bar{\theta}\right)\cos\left(\bar{\theta}\right),  \\
\epsilon^{\prime} &= -\frac{(m \nu)^{\prime}}{m\nu} \cos\left(2\bar{\theta}\right).
\end{align}
The system is not yet fully described, since we have only the dynamics
for a composition of $I$ and $J$ through $\epsilon$. Introducing $\gamma$
by the relation
\begin{equation}
\e^{\gamma} = \sqrt{\frac{I}{J}},
\end{equation}
we easily obtain
\begin{equation}
 \gamma^{\prime} = - 2\frac{(m \nu)^{\prime}}{m \nu} \frac{\sin
\left(\bar{\theta}\right)\cos
\left(\bar{\theta}\right)}{\cosh \left( \epsilon \right)},
\end{equation}
and $I $ and $J$ can be recovered using
\begin{align}
 I &= \e ^{\gamma} \cosh( \epsilon), \label{rec_I}\\
 J &= \e ^{- \gamma} \cosh( \epsilon) \label{rec_J}.
\end{align}
Finally the complete set of equations that replaces Eqs.~\eqref{eom_I},
\eqref{eom_theta}, \eqref{eom_J}, \eqref{eom_psi} already accounting the
constraint \eqref{norm_cond} is
\begin{align}
 \bar{\theta}^{\prime} &= \nu + \frac{(m \nu)^{\prime}}{m \nu} \tanh \left(\epsilon\right) \sin \left( \bar{\theta} \right)\cos \left( \bar{\theta} \right) , \\
\epsilon^{\prime} &= -  \frac{(m \nu)^{\prime}}{m \nu} \cos \left(2
\bar{\theta}\right), \\
 \gamma^{\prime} &= - 2\frac{(m \nu)^{\prime}}{m \nu} \frac{\sin
\left(\bar{\theta}\right)\cos
\left(\bar{\theta}\right)}{\cosh \left( \epsilon \right)}.
\end{align}
Using these new variables, the two real linearly independent solutions
can be recast as
\begin{align}
\tzeta_k^{a} &= \frac{\e^{\gamma/2}}{\sqrt{m \nu}} \left[
\e^{\epsilon/2} \sin\left(\bar{\theta}\right) - \e^{-\epsilon/2} \cos\left(\bar\theta\right)\right], \\
\tzeta_k^{b} &= \frac{\e^{-\gamma/2}}{\sqrt{m \nu}} \left[
\e^{\epsilon/2} \sin\left(\bar{\theta}\right) + \e^{-\epsilon/2} \cos\left(\bar\theta\right)\right].
\end{align}

Now we have to rewrite the adiabatic vacuum initial conditions, for that
purpose we use the adiabatic limit ${(m\nu)^{\prime}}/{m\nu} \rightarrow
0$. Expanding in the leading order in ${(m\nu)^{\prime}}/{m\nu}$ the set
of equations provides
\begin{align}
\epsilon &\approx \epsilon_0, \\
\gamma &\approx \gamma_0, \\
\bar\theta &\approx \bar\theta_0 + k\eta.
\end{align}
Using the above approximations to calculate the complex $\tzeta_k$, we
have, as an initial condition, that the following choice recovers the
leading order WKB approximation, Eqs.~\eqref{ini_vacuo_1} and
\eqref{ini_vacuo_2}:
\begin{equation}
 \epsilon_0 = \gamma_0 = 0,
\end{equation}
which, naturally, satisfies the Eq.~\eqref{norm_cond}. The real solutions using this choice are
\begin{align*}
\tzeta_k^{a} &= \sqrt{\frac{2}{m \nu}} \sin\left(\bar{\theta}-\frac{\pi}{4}\right), \\
\tzeta_k^{b} &= \sqrt{\frac{2}{m \nu}} \cos\left(\bar\theta-\frac{\pi}{4}\right),
\end{align*}
then, consequently, the complex solution is
\begin{equation*}
\tzeta_k = \frac{\e^{-i\left(\bar{\theta}-{\pi}/{4}\right)}}{\sqrt{2m \nu}}.
\end{equation*}
Because it is just a phase, we can choose $\bar\theta_0 = \pi/4$. 

The same treatment follows through in the case of the dimensionless tensor
perturbation, with the only difference being the ``mass'' definition
\begin{equation}\label{massh}
m_h = 4\kappa^2R_Hz_h^2 = \frac{a^3R_H}{N},\qquad \nu_h = \frac{N k}{a R_H}.
\end{equation}

A general purpose integrator for systems described by an action in
the form~\eqref{quad_hamilt} was implemented as part of the
\textsf{NumCosmo} (numerical cosmology library)~\cite{Vitenti2012c}. The
abstract implementation of the harmonic oscillator action angle is
provided by the object \textsf{NcmHOAA}~\cite{Vitenti2017}. The adiabatic
and tensor mode objects, respectively \textsf{NcHIPertAdiab} and
\textsf{NcHIPertGW}, are built on top of \textsf{NcmHOAA}, connecting it
to the exponential potential background model \textsf{NcHICosmoVexp}. We
should stress that these codes can be used to any background model, all
one needs to do is to implement the interface which calculates the mass
$m$ and frequency $\nu$ of the model.

\begin{table}[ht]
\begin{tabular}{|c|c|c|c|c|}
\hline
&$d$ & $\sigma$ & $\alpha_{\mathrm{b}}$ & $\mathcal{X}_ {\mathrm{b}}$ \\
\hline
&&&&\\[-0.8em]
set1 &$-9\times10^{-4}$ & $9$ & $8.3163 \times 10^{-2}$ & $2 \times 10^{36}$
\\
\hline
&&&&\\[-0.8em]
set2 &$-9\times10^{-4}$ & $100$ & $7.4847\times10^{-3}$ & $4 \times 10^{36}$
\\
\hline
&&&&\\[-0.8em]
set3 &$-0.1$ & $4$ & $10^{-5}$ & $6 \times 10^{37}$
\\
\hline
&&&&\\[-0.8em]
set4 &$-0.1$ & $4$ & $10^{-7}$ & $6 \times 10^{37}$
\\
\hline
\end{tabular}
\caption{Model parameters for four different cases in which the present
model produces $\Delta_{\zeta}$ close to $10^{-10}$, and scale invariant
spectra. The relevant background quantities are presented in
Fig.~(\ref{Fig:evol_x}) through (\ref{Fig:LRicci}), while the modes
evolution can be seen in Fig.~(\ref{Fig:evol_zeta_h}). The DE scale is
fixed at $\Omega_\Lambda = 1$.} \label{tab_par_1}
\end{table}

\section{Numerical solutions for the perturbations}
\label{sec_sol_pert}

Using the AA variables, we can calculate the power spectra for both
scalar and tensor perturbations. In this section, we will present these
solutions, and discuss how the basic parameters of the model influence
the primordial perturbations. We will focus in case \B, which is a
complete background that addresses the problem of DE in bounce models by
means of a single scalar field, in this analysis we keep fixed both
$\Omega_\Lambda = 1$ and $\Omega_d = 1$. Here we analyze in detail four
parameter sets defined in Tab.~\ref{tab_par_1}.

\begin{figure}[ht]
\begin{center}
\includegraphics[width=0.5\textwidth]{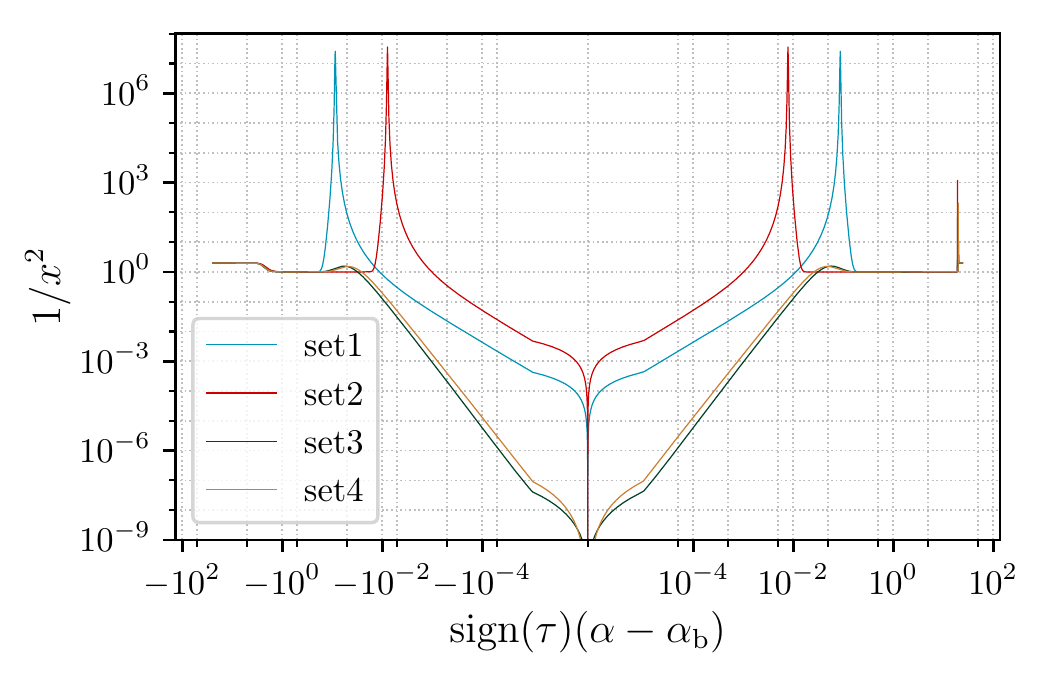}
\end{center}
\caption{Evolution of the $1/x^2$ term given four sets of parameters,
set1, set2, set3 and set4, appearing in Tab.~\ref{tab_par_1}. The
distance between peaks is controlled by the $\sigma$ parameter, while its
height is determined by how much the background bouncing trajectory gets
close to the cyclic solutions presented in
Fig.~(\ref{Fig:phase_quantum}), which in turn is defined by the value of
$\alphab$. } \label{Fig:evol_x}
\end{figure}

Before moving to the solutions themselves, it is worth to take a look at
the differences between the scalar mode and the tensor mode evolution.
The super Hubble solutions for Eqs.~\eqref{quad_hamilt} are, at leading
order,
\begin{align}\label{solHa}
\tzeta_k &= A^1_k\left[1 + \mathcal{O}\left(\nu^2\right)\right] + A^2_k\left[ \int\frac{\dd\tau}{m} + \mathcal{O}\left(\nu^2\right)\right], \\
\thg_k &= B^1_k\left[1 + \mathcal{O}\left(\nu_h^2\right)\right] + B^2_k\left[ \int\frac{\dd\tau}{m_h} + \mathcal{O}\left(\nu_h^2\right)\right],
\end{align}
where changing the integration limits in the integrals above result in a
redefinition of $A^1_k$ and $B^1_k$. Using the expression for the masses in
Eqs.~\eqref{masszeta} and \eqref{massh}, we recast the integral as
\begin{align}\label{eq:amps}
\int\frac{\dd\tau}{m} &= \frac{1}{R_H}\int\dd\tau \frac{N}{x^2a^3} = \frac{1}{R_H}\int \frac{\dd{}t}{x^2a^3}, \\ \label{eq:amph}
\int\frac{\dd\tau}{m_h} &= \frac{1}{R_H}\int\dd\tau \frac{N}{a^3} = \frac{1}{R_H}\int \frac{\dd{}t}{a^3}.
\end{align}
In the classical contracting branch of case \B, $x$ varies between
$(1/\sqrt{2},\; 1)$, while the scale factor goes through a large
contraction. In other words, the value of this integral will be dominated
by the values of $a$ near the bounce phase, where $a$ attains its
smallest value (see~\cite{Vitenti2012} for a detailed analysis of this
integral). Nonetheless, when the quantum phase begins, the value of $x$
is no longer restricted to $(1/\sqrt{2},\; 1)$. For example, in
Fig.~\ref{Fig:evol_x}, we show three time evolutions for $1/x^2$ using
four different sets of parameters.

\begin{figure}[ht]
\begin{center}
\includegraphics[width=0.5\textwidth]{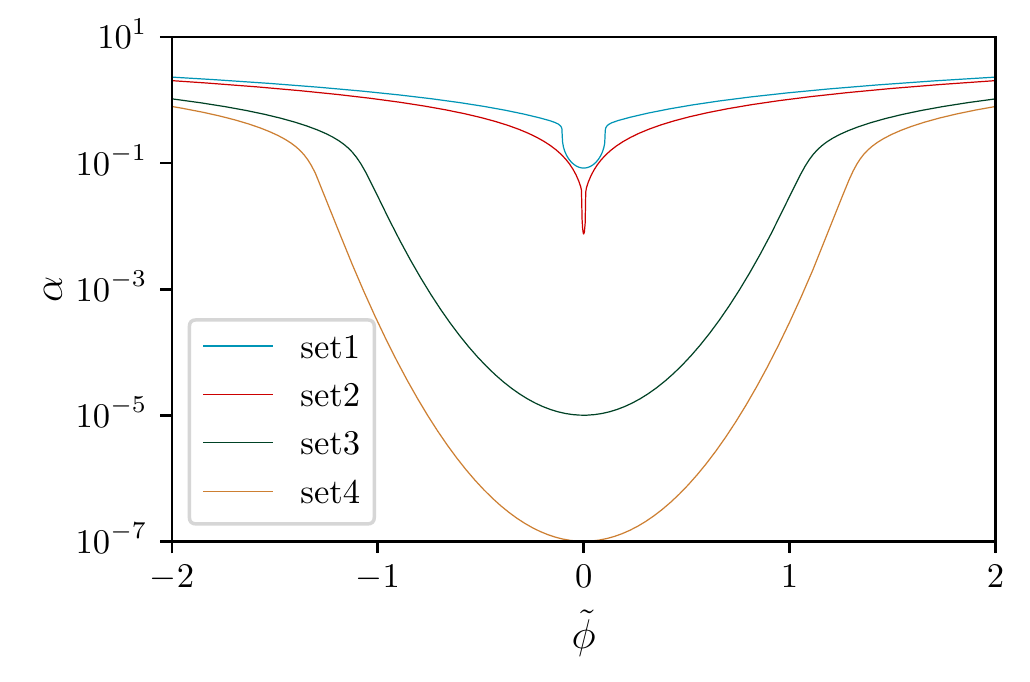}
\end{center}
\caption{Phase space evolution for the four sets of parameters appearing
in Tab.~\ref{tab_par_1}. Note that the set1 and set2 curves are almost
vertical near the bounce. This happens because they pass close to the
periodic trajectories (see Fig.~\ref{Fig:phase_quantum} for a full
picture of the phase space trajectories). At these points, $x \propto
\dd\phi/\dd\alpha \approx 0$, which results in the peaks seen in
Fig.~\ref{Fig:evol_x}. Contrastingly, the set3 and set4 curves pass far
from the center points of Fig.~\ref{Fig:phase_quantum}, resulting in a
smoother transition through the bounce phase. With all parameters fixed,
we can control how close one gets to the cyclic solutions by increasing
the value of $\alphab$. One can also see in the figure, by comparing set1
with set2, that a larger $\sigma$ induces a faster bounce.}
\label{Fig:evol_phi_alpha}
\end{figure}

During the matter phase $1 / x^2 \approx 2$, and during the stiff matter
domination, $1/x^2 \approx 1$. Therefore, in the classical phase, the
presence of $1/x^2$ in the integral~\eqref{solHa} increases its value by
a maximum factor of two. On the other hand, throughout the quantum phase,
different parameter sets result in quite different behaviors, as can be
seen in Fig.~\ref{Fig:evol_x}. The set1 curve shows that the presence of
$1/x^2$ in the aforementioned integral will result in a sharp increase in
the spectrum amplitude around $\vert\alpha-\alphab\vert \approx 10^{-1}$.
This effect takes place slightly closer to the bounce in set2.
Furthermore, in set3 we show a solution where the peaks are negligible,
and hence there is no further increase of the perturbation amplitudes
around the bounce. The phase space plot in Fig.~\ref{Fig:evol_phi_alpha}
elucidates what is happening. The set1 and set2 configurations are such
that the system passes closes to the cyclic solutions (see
Fig.~\ref{Fig:phase_quantum}), and the curve is near vertical, $\alpha$
changes abruptly with $\phi$, making $x$ close to zero during this
interval. The scalar field shortly behaves as a DE fluid, implying a
momentarily large deceleration (acceleration), which enhance the scalar
perturbations. We also added the set3 and set4 to show solutions which
pass far from these cyclic solutions. In this case, the evolution of
$\alpha$ with respect to $\phi$ is smoother, resulting in bigger values
of $x$.

\begin{figure*}[ht]
\begin{center}
\includegraphics[width=0.5\textwidth]{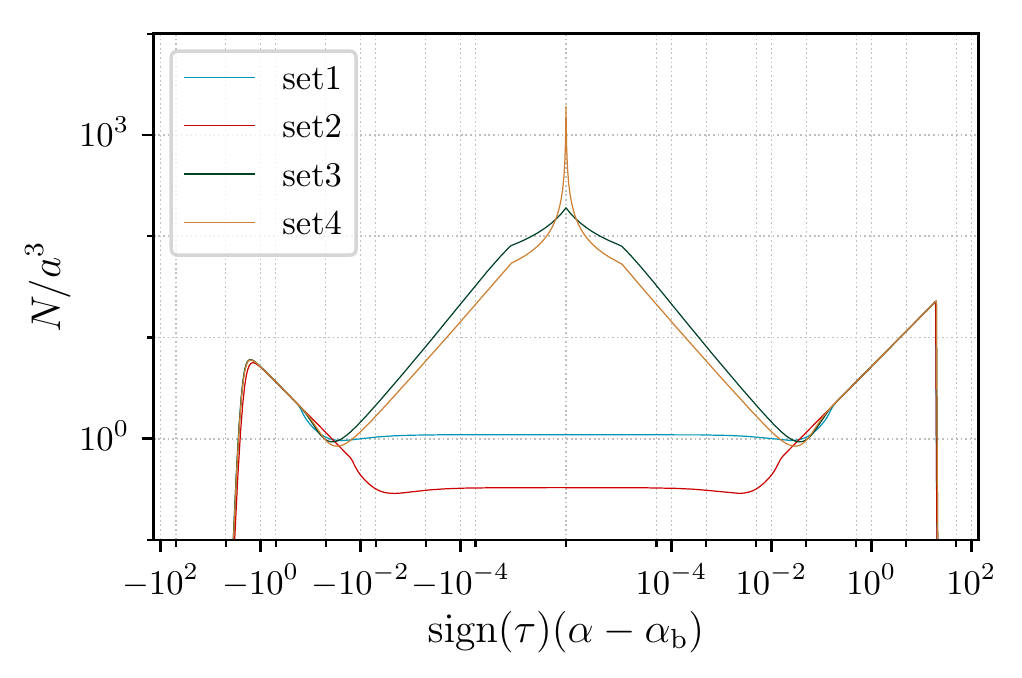}\includegraphics[width=0.5\textwidth]{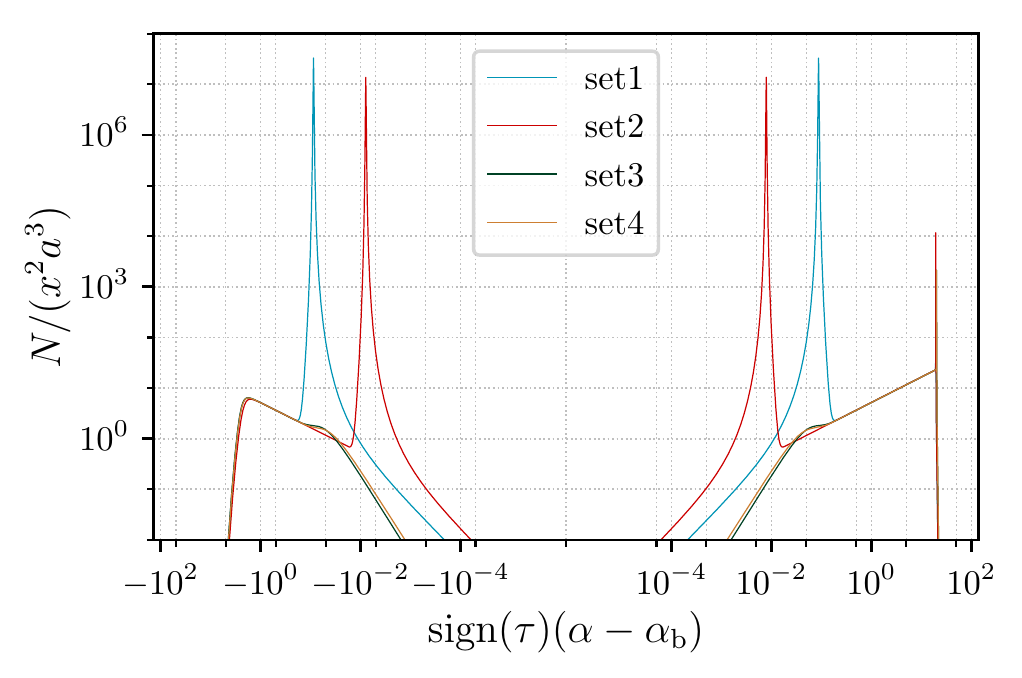}
\end{center}
\caption{Integrand of the super Hubble approximation for the tensor modes
(left panel) and scalar modes (right panel). It is worth noting that the
presence of the $1/x^2$ term in the scalar mode integrals, which goes
through zero during the bounce phase, overcomes any possible additional
contribution to the amplitude from the peak in the lapse function $N =
\tau/H$. Nevertheless, the tensor modes are sensitive to the peaks in
$N$.} \label{Fig:integ}
\end{figure*}

On the other hand, the tensor mode amplitudes do not depend on $x$, see
Eq.~\eqref{eq:amph}. In Fig.~\ref{Fig:integ}, we present the evolution of
both integrands of Eqs.~\eqref{eq:amps} and \eqref{eq:amph}. Note that
the first peak to the left in both figures is just the usual increase in
amplitude related to the contraction, while during the bounce itself we
have two different behaviors depicted. The tensor modes, which depend
only on the lapse function $N$ and the scale factor, are sensitive to the
peaks of $N$ at the bounce, whereas for scalar modes, the presence of the
$1/x^{2}$ term in the integrand overcomes the $N$ dependence around the
bounce. Here we would like to emphasize that the exact dynamics
controlling the bounce is extremely important to determine the amplitude
of the spectrum. Integrating only the classical phase, the first left
peak in both integrands would provide similar amplitudes for both scalar
and tensor modes, implying in a larger than one tensor-to-scalar ratio
(see, for example, Ref.~\cite{Quintin2015}). However, one cannot stop at
the classical phase, since the bounce evolution will left a definite
imprint on the amplitudes: an increase in the tensor mode amplitude at
the bounce, and two amplifications of the scalar modes at two symmetric
points around the bounce. Hence, any new physics around the bounce
producing this kind of effect can be physically relevant, and its
consequences must be evaluated with care.

The contracting phase with a matter era puts this model in the category
of the matter bounce scenarios. Previous works on the field obtained the
bounce by means of a second scalar field with a ghost-type Lagrangian.
Choosing wisely the parameters of the ghost scalar field, the bounce
takes place only very close to the singularity, and the perturbations are
studied, sometimes, without taking it into account. The results obtained
in the literature about matter bounces can be summarized as follows: the
spectrum is scale invariant; the tensor-to-scalar ratio is usually larger
then measured in CMB if the scalar field is canonical and the bounce is
symmetric; attempts to solve this problem assuming the validity of GR all
along results in the increase of non-Gaussianities, which seems to
suggest a no-go theorem for bounce cosmologies~\cite{Quintin2015,
Li2017}. Our results point to a new direction: bounces which are out of
the scope of GR can lead to new ways to decrease $r$, and they should be
investigated with care. In our model, the decrease of the
tensor-to-scalar ratio relies on the actual bounce dynamics in a
non-trivial way, due to quantum effects. The increase in amplitude of the
scalar and tensor modes take place at different times and are controlled
by distinct parameters. The consequences of that for non-Gaussianities is
something yet to be investigated.

\begin{figure}[ht]
\begin{center}
\includegraphics[width=0.5\textwidth]{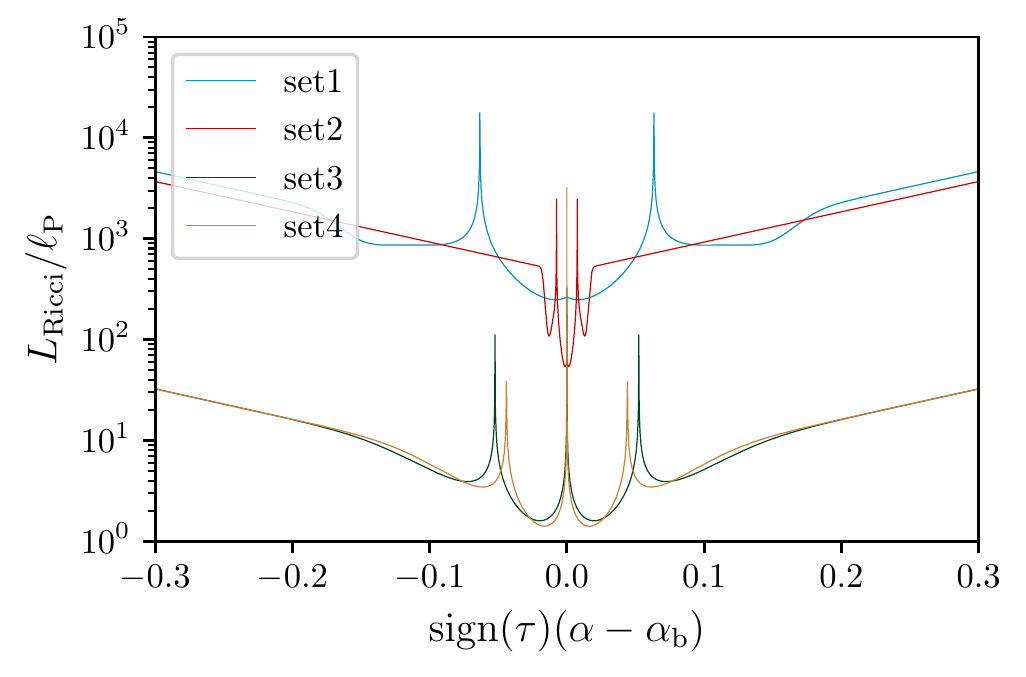}
\end{center}
\caption{Time evolution of the Ricci scale for all sets appearing in
Tab.~\ref{tab_par_1}. The parameter $d$ controls how close the scale gets
to the Planck length and set3 and set4 are in the limit of validity of our
model. Thus, a large (in module) value of $d$ would violate this
constraint. Note also that faster bounces (for instance, set2) result in
stronger oscillations of $L_R$ near the bounce. This means that faster
bounces must take place at even higher scales in order to avoid a
violation of $L_R/\lP > 1$ during the oscillations.} \label{Fig:LRicci}
\end{figure}

Concerning the amplitude growth in the classical regime, a known result
is that they grow more substantially during a matter epoch then during
the stiff matter one. This can be seen by looking at the super Hubble
approximations in Eqs.~\eqref{eq:amps} and \eqref{eq:amph}. During the
matter domination, $N/a^3 \approx \tau / a^{3/2}$, while for stiff matter
$N/a^3 \approx \tau$, where we are using that $N = \tau/H$, and $H
\propto a^{-3/2}$ in the matter phase and $H \propto a^{-3}$ at the stiff
phase. Since this part of the amplitude growth is determined by the
matter epoch duration, it is closely connected with the parameters $d$
and $\Xb$, which also control the bounce depth.

\begin{figure*}[ht]
\begin{center}
\includegraphics[width=0.5\textwidth]{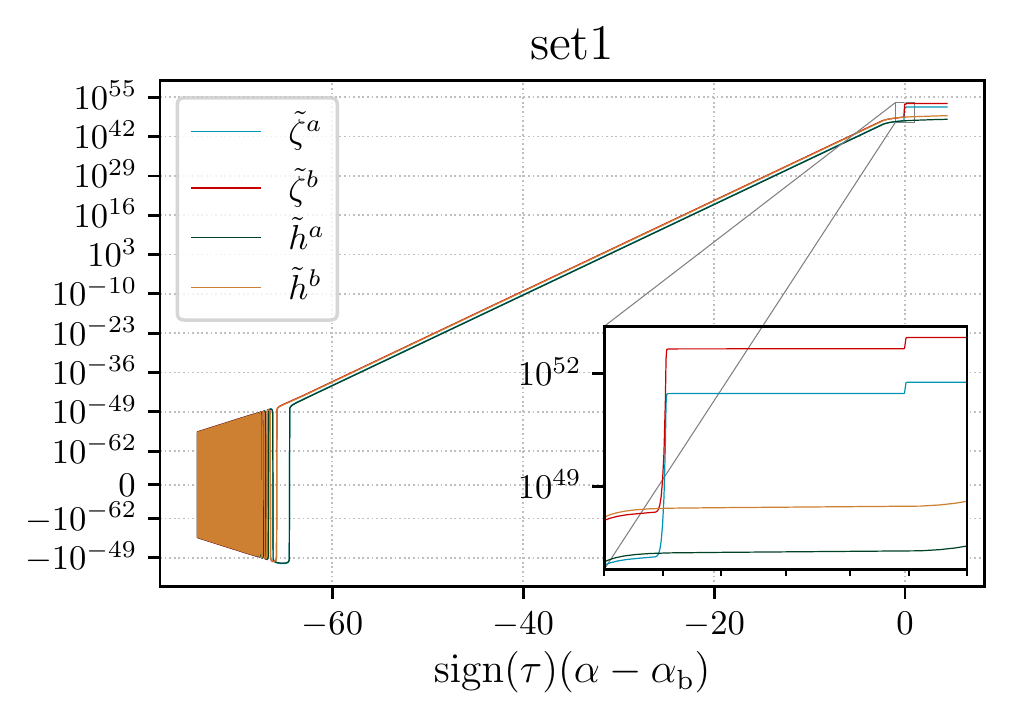}\includegraphics[width=0.5\textwidth]{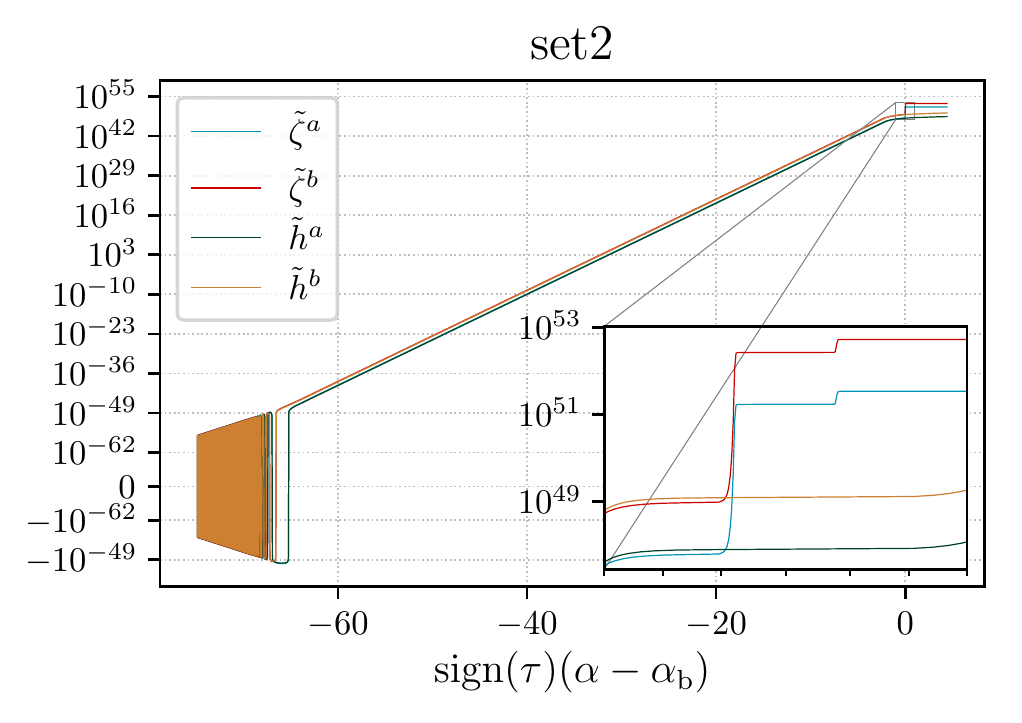} \\
\includegraphics[width=0.5\textwidth]{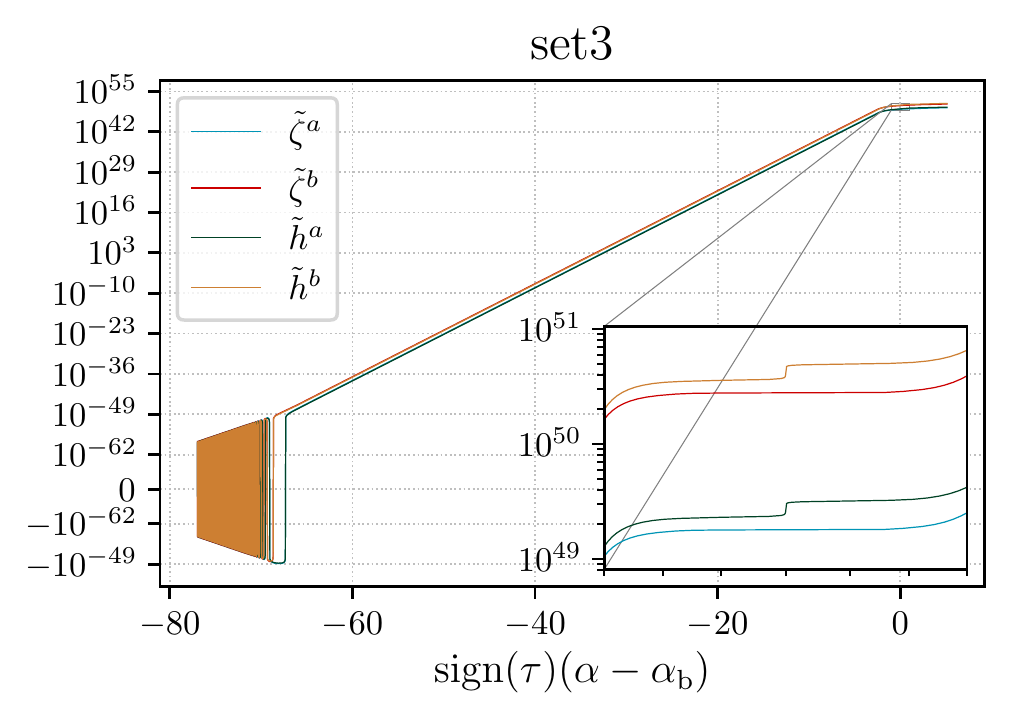}\includegraphics[width=0.5\textwidth]{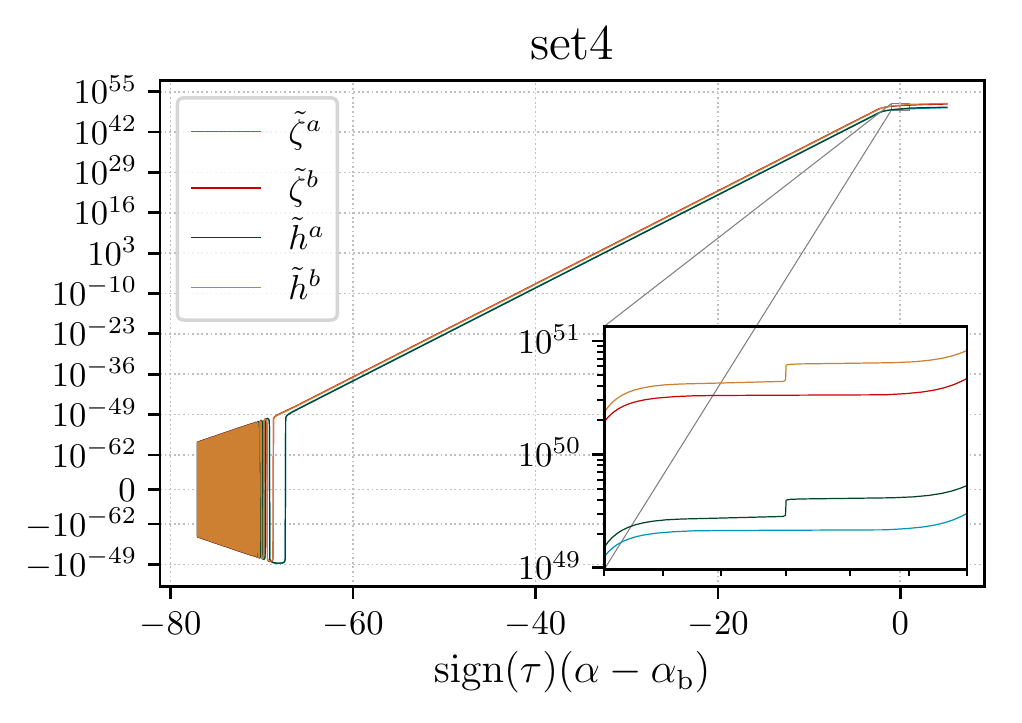}
\end{center}
\caption{Evolution of the mode functions $\tzeta_k$ and $\thg_k$ for
set1, set2, set3 and set4. In the detail, we can see the result of the
integration of the two peaks in Fig.~\ref{Fig:evol_x} for the scalar mode
in the upper panels and the single peak of Fig.~\ref{Fig:integ}
integration for the tensor mode in the lower panels. For example, in the
upper left figure, the first peak around $-0.1$ increases the amplitude
of $\tzeta^a$ and the second peak at $+0.1$ double this value. In
contrast, since the tensor perturbations amplitude does not depend
directly on the evolution of $x$, they are not modified by these peaks.
Nonetheless, the tensor amplitude is sensible to peaks in the lapse
function $N$. Hence, for set3 and set4 where these peaks are pronounced,
we have an increase in the amplitude of tensor perturbations at the
bounce, which is otherwise overcome by scalar perturbations in the cases
where the $1/x^2$ term become relevant.} \label{Fig:evol_zeta_h}
\end{figure*}

We emphasize that the parameter choices are implicit determinations of
the background model initial conditions (including the wave function
parameters), as already mentioned. The results for the power spectra at
the pivotal mode $k_*$ are:
\begin{align*}
&\mathrm{set1:} &\left.\Delta_{\zeta_k}\right\vert_{k=k_*} = 1.4 \times 10^{-10},\; &r = 1.9 \times 10^{-7}, \\
&\mathrm{set2:} &\left.\Delta_{\zeta_k}\right\vert_{k=k_*} = 4.6 \times 10^{-11},\; &r = 1.3 \times 10^{-5}, \\
&\mathrm{set3:} &\left.\Delta_{\zeta_k}\right\vert_{k=k_*} = 1.2 \times 10^{-14},\; &r = 56, \\
&\mathrm{set4:} &\left.\Delta_{\zeta_k}\right\vert_{k=k_*} = 1.7 \times 10^{-14},\; &r = 59.
\end{align*}
The time evolution for this same pivotal mode is shown in
Fig.~\ref{Fig:evol_zeta_h}. Observe that, for set1 and set2, the extra
enhancement of the scalar amplitude due to the quantum effects takes it
to a value close to the observed one
$\left.\Delta_{\zeta_k}\right\vert_{k=k_*} \approx 10^{-10}$. On the
contrary, the power spectra obtained from set3 and set4 have an amplitude
smaller than the required by observations, even though their bounces are
deeper ($\Xb \approx 10^{36}$ for set1 and set2 and $\Xb \approx 10^{37}$
for set3 and set4). In principle, one could choose the parameters in
order to make the bounce deeper, hoping to get the right amplitude.
Nevertheless, we must take care to not go beyond the scale of validity of
these models. One should verify whether the energy scale of the bounce is
not dangerously close to the Planck energy scale, where our simple
approach would not be appropriate. The curvature scale at the bounce is
given by the Ricci scalar,
\begin{equation}
R = 12 H^2 + 6\dot{H}, \quad L_R = 1/\sqrt{R},
\label{ricci}
\end{equation}
and Ricci scale $L_R$ should not be smaller than the Planck length.
Figure~\ref{Fig:LRicci} displays the Ricci scale evolution for all
parameter sets. This figure shows that the absolute value of $d$ controls
the minimum scale $L_R$ attained around the bounce. This means that we
could not increase the amplitudes of set3 and set4 by increasing $\vert
d\vert$ without violating the validity of our approach.

In what concerns the spectral index, our modes of interest cross the
potential during the matter domination phase. As such, their spectrum are
very close to scale invariant. Again, this can be changed using a slight
negative value for the matter phase EoS.

\section{Conclusion}
\label{sec_conclusion}

We have studied the evolution of cosmological perturbations of quantum
mechanical origin in a nonsingular cosmological model containing a single
scalar field with exponential potential. The bounce is driven by quantum
corrections of gravity in high-energy scales, but still smaller than the
Planck energy scale, in which the canonical quantization of gravity may
be applied.

In Secs.~\ref{sec_back}, \ref{sec_quant} and \ref{sec_matc} we presented
two possibilities for the homogeneous and isotropic background dynamics:
case \A, which contains a DE epoch only in the contracting phase, and
case \B, in which the DE epoch happens only in the expanding phase. In
both cases, the scalar field behaves like a dust fluid in the asymptotic
past and future. The free parameters at our disposal were mapped to
quantities with physical significance, as the duration of the matter
contraction, the energy scales of the DE epoch and of the bounce phase,
and their selection was equivalent to choose initial conditions for the
numerical integration, hence allowing a better physical control of the
whole scenario.

We restricted our attention mainly to case \B, which is a matter bounce
model with a DE epoch in the expanding phase, offering a complete
background solution in which the DE epoch arises naturally in the
expanding phase by means of the same scalar field that drives the bounce
and the matter contraction. This is a bouncing model with DE where its
presence does not cause any trouble in defining adiabatic vacuum initial
conditions in the far past of the contracting phase (as described in
Ref.~\cite{Maier2011}), because it is dominated by dust. We numerically
calculated the evolution of cosmological perturbations of quantum
mechanical origin in such backgrounds. Scalar and tensor perturbations
result to be almost scale invariant, and the parameters of the background
can be adjusted to yield the good amplitudes for scalar and tensor
perturbations, producing $r < 0.1$. We have also seen that it was in the
quantum bounce that the scalar perturbations were amplified with respect
to tensor perturbations. Hence, the same quantum effects which produces
the background bounce do also induce the property $r < 0.1$. Our result
shows that when GR is violated around the bounce, the influence of this
phase on the evolution of cosmological perturbations can be nontrivial,
and must be evaluated with care. Since we have only one scalar field, the
perturbations were solved numerically for the whole background history,
without approximations and matching conditions. Using the AA variables,
we were able to construct a robust code where the details of the bounce
effects on the perturbations could be appreciated for different
background features. We have also found that longer the dust contraction,
bigger are the amplitudes, a fact that was not noticed in other
investigations.

We have thus obtained a bouncing model with a single canonical scalar
field which produces the observed features of cosmological perturbations
at linear order. Usually, canonical scalar fields in the framework of GR
produces $r \geq 1$ for symmetric bounces ~\cite{Allen2004,
Battarra2014}. We should emphasize that, in our model, we get $r < 1$ not
because it is asymmetric, but because of violations of GR around the
bounce due to quantum effects. The next step should be to evaluate
non-Gaussianities in this model. Note that, as long as GR is not
satisfied around the bounce, arguments based on the full validity of GR
even at the bounce suggesting that non-Gaussianities should be huge in
single scalar field bouncing models do not apply~\cite{Quintin2015,
Li2017}. The evaluation of the non-Gaussianities must be made with care
through the quantum bounce, as long as GR is not valid there. A
consistent framework must be developed in order to perform this
calculation correctly. This is one of our future investigations.

Note that this is a very simple model, with a single scalar field, but it
is astonishing that it can produce alone the right amount of cosmological
perturbations, and also yield a future DE phase. Hence, it is quite
reasonable to pursue this route and try to complete the model in order to
obtain more accurate scenarios for the real Universe. One possibility is
to perform the same calculations when other fluids are present, as the
classical extension of this model presented in Ref.~\cite{Copeland1998}.
This is a much more involved calculation, where entropy perturbations
must be considered.

In a future work, we will also study more deeply case \A, which is a very
interesting theoretical laboratory to investigate more precisely the
influence of DE in a contracting phase. In that scenario, the
cosmological perturbations evolve in a situation where usual adiabatic
vacuum initial conditions can be posed normally, since DE behavior is
transient and the background is matter dominated in the far past.

Finally, let us emphasize that the present model can accommodate non
negligible primordial gravitational waves which might be detected in the
near future. Hence, these models are observationally testable.

\acknowledgments

APB and NPN would like to thank CNPq of Brazil for financial support.
SDPV acknowledges the financial support from a PCI postdoctoral
fellowship from Centro Brasileiro de Pesquisas F\'{i}sicas of Brazil,
and from BELSPO non-EU postdoctoral fellowship.

\bibliographystyle{apsrev}
\bibliography{references}

\end{document}